\documentclass[A_4paper,11pt,final]{article}
\pdfoutput=1
\usepackage[utf8]{inputenc}
\usepackage{jheppub}
\usepackage{amsmath}
\usepackage{braket}
\usepackage{float}
\usepackage{tabularx}
\usepackage{tabulary}
\usepackage{multirow}
\usepackage[inline]{showlabels}
\usepackage{tikz}
\usetikzlibrary{positioning}
\usetikzlibrary {arrows.meta,bending,positioning}

\usepackage{subcaption}

\newcommand*\diff{\mathrm{d}}

\usepackage{xspace}
\newcommand*{\ie}{i.e., }
\newcommand*{\eg}{e.g., }
\newcommand*{\fig}{fig.\@\xspace}
\newcommand*{\figs}{figs.\@\xspace}
\newcommand*{\eq}{eq.\@\xspace}
\newcommand*{\eqs}{eqs.\@\xspace}
\newcommand*{\cf}{cf.\@\xspace}

\title{The Timescales of Quantum Breaking}

\author[a]{Marco Michel,}
\author[b]{Sebastian Zell}

\affiliation[a]{Department of Physics, Ben-Gurion University of the Negev, Beer-Sheva 84105, Israel}
\affiliation[b]{Centre for Cosmology, Particle Physics and Phenomenology -- CP3,
	Universit\'e catholique de Louvain, B-1348 Louvain-la-Neuve, Belgium}
\emailAdd{michelma@post.bgu.ac.il}
\emailAdd{sebastian.zell@uclouvain.be}

\abstract{Due to the inevitable existence of quantum effects, a classical description generically breaks down after a finite quantum break-time $t_q$. We aim to find criteria for determining $t_q$. To this end, we construct a new prototype model that features numerous dynamically accessible quantum modes. Using explicit numerical time evolution, we establish how $t_q$ depends on the parameters of the system such as its particle number $N$. The presence of a classical instability leads to $t_q\sim\ln N$ or $t_q\sim \sqrt{N}$. In the stable case, we observe $t_q\sim N$, although full quantum breaking may not take place at all. We find that the different regimes merge smoothly with $t_q\sim N^\gamma$ ($0<\gamma<1$).
As an outlook, we point out possibilities for transferring our results to black holes and expanding spacetimes.}

\date{}

\makeatletter
\gdef\@fpheader{\phantom{text}}
\makeatother

\begin{document}
	
\maketitle
	
\section{Breakdown of the classical description}
	
\subsection{Computing the quantum break-time}
\begin{figure}[h]
	\centering 
\begin{tikzpicture}
	
\node [draw,
minimum width=4cm,
]  (initial) {$\ket{\phi(t=0)}$};

\node [draw,
minimum width=4cm, 
right=5cm of initial
] (quantum) {$\ket{\phi(t=t_f)}$};

\node [draw, 
minimum width=4cm, 
below = 2cm of initial
]  (classicalL) {$\braket{\phi(t=0)|\hat{\Psi}|\phi(t=0)}$};

\node [draw, 
minimum width=4cm, 
below = 2cm of quantum
]  (expectation) {$\braket{\phi(t=t_f)|\hat{\Psi}|\phi(t=t_f)}$};

\draw[-{Stealth[width=3mm]}] (initial.east) -- (quantum.west)
node[midway,above]{Quantum time evolution};

\draw[-{Stealth[width=3mm]},dashed] (classicalL.east) -- (expectation.west)
node[midway,above]{Classical time evolution}
node[midway,below]{???};

\draw[-{Stealth[width=3mm]}] (initial.south) -- (classicalL.north)
node[midway, left, align=left]{Classical \\limit};

\draw[-{Stealth[width=3mm]}] (quantum.south) -- (expectation.north)
node[midway, right, align=left]{Expectation \\value};
\end{tikzpicture}

\caption{Schematic depiction of quantum breaking. Starting from an initial state $\ket{\Phi(t=0)}$, exact dynamics corresponds to computing time evolution of the quantum state and then evaluating its expectation value. In the classical approximation, one already takes the expectation value of the initial state and then calculates time evolution classically. Quantum breaking occurs when these two approaches differ significantly.}
\label{fig:breaking}
\end{figure}
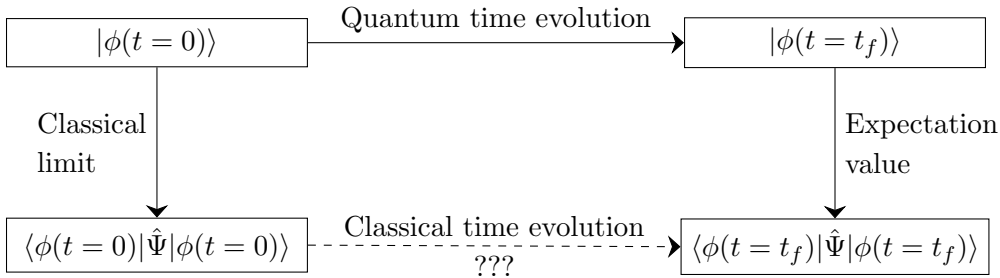

A fundamental description of our world must include quantum effects. Nevertheless, classical theories yield highly accurate results in many situations. Why this is the case and how localized wave functions correspond to classical trajectories have been  important questions since the early days of quantum mechanics \cite{Einstein1917,DeBroglie1926,Brillouin1926,Schroedinger1926,Debye1926,Heisenberg1927,Kennard1927,Ehrenfest1927}.\footnote
{See \cite{Einstein1917E} for an English translation of \cite{Einstein1917}.}
In particular, the classical solution is exact in special circumstances \cite{Ehrenfest1927}. Generically, however, the classical description only has a limited range of applicability. Starting from a given initial state, the solution computed in the classical approximation cannot track indefinitely the expectation value derived from the true quantum evolution (see \fig \ref{fig:breaking}). Following \cite{Dvali:2013vxa}, we shall refer to the timescale $t_q$ after which quantum effects invalidate the classical description as the \textit{quantum break-time}. 

An explicit study of $t_q$ was first performed in small systems corresponding to a single quantum oscillator \cite{Berman1978, Berry1979,Berry19792}. In certain situation, a fast quantum breaking was found, corresponding to the scaling \cite{Berman1978, Berry1979}\footnote
{See \cite{Balazs1987,Balazs1989, Saraceno1990} for further investigations leading to \eq \eqref{logarithmicScaling}.}
\begin{equation} \label{logarithmicScaling}
	t_q \sim \ln \hbar^{-1} \;,
\end{equation}
where $\hbar$ is the Planck constant.
If the properties of the system are altered, the quantum break-time can also be much longer \cite{Berry1979}:
\begin{equation} \label{linearScaling}
	t_q \sim  \hbar^{-1} \;.
\end{equation}
Finally, it was pointed out that also intermediate scalings are possible, \cite{Berry19792,OConnor1991,Sepulveda1992,Tomsovic1993}:
\begin{equation} \label{fractionalScaling}
	t_q \sim  \hbar^{-\gamma} \;,
\end{equation}
where $0<\gamma<1$. 
The timescales \eqref{logarithmicScaling} and \eqref{fractionalScaling} that are faster than linear emerge when an instability leads to chaotic behavior  \cite{Chirkov1981,Chirkov1987}.

Quantum effects can also play a crucial role in systems of many particles -- famous examples are Bose-Einstein condensation \cite{Bose1924,einstein1925} and superconductivity \cite{ginzburg1950, Bardeen1957}. Nevertheless, one might expect that a classical approximation becomes increasingly accurate on macroscopic scales. Intriguing indications against this common belief come from gravity. As detailed below, a microscopic model of a black hole -- the \textit{quantum N-portrait} \cite{Dvali:2011aa} -- made  evident that crucial deviations from the classical description can also occur for  arbitrarily large objects \cite{Dvali:2012rt,Dvali:2012en,Dvali:2012wq,Flassig:2012re,Dvali:2013vxa}.

This sparked an interest in analogous phenomena in non-relativistic quantum many-body systems and a one-dimensional Bose gas with periodic boundary conditions \cite{LiebLiniger1963,Kanamoto2002} was employed as prototype model. This system, to which we shall refer as \textit{periodic prototype model (PPM)}, features an attractive contact interaction of strength $\alpha$ and exhibits two different regimes which are separated by a second order quantum phase transition \cite{Kanamoto2002} (see also \cite{Kanamoto2004, Kanamoto2005,Sakman2005,Sykes2007, kanamoto2009, kanamoto2010}). The relevant control parameter is the collective coupling $\lambda = \alpha N$, where $N$ is the particle number. One can normalize $\lambda$ such that for $\lambda <1$ the system is undercritical and the ground state is homogeneous while $\lambda>1$ leads to overcriticality and a bright soliton.  
 For $\lambda>1$, the PPM exhibits quantum breaking on the timescale \cite{Dvali:2013vxa}
\begin{equation} \label{logarithmicScalingN}
	t_q \sim  \frac{1}{\sqrt{\lambda - 1}} \ln N \;.
\end{equation}
 As in \cite{Chirkov1981,Chirkov1987}, this logarithmic quantum break-time emerges due to an instability, where the prefactor in \eq  \eqref{logarithmicScalingN} is set by the classical Lyapunov exponent \cite{Dvali:2013vxa}. 

Among various further studies \cite{Flassig:2012re, Flassig:2015nba,Dvali:2015ywa,Dvali:2015wca,Panchenko:2015dca,Dvali:2016zqx, Piroli:2016jdu,Dvali:2017ruz,Hummel:2018brt,Rautenberg:2019kmd, Zell:2019vpe} about quantum effects in the PPM, an important question concerns the behavior of the system at the critical point $\lambda =1$. In this case, there are indications that deviations from a classical description occur on the timescale \cite{Dvali:2015wca}:
\begin{equation} \label{rootScalingN}
	t_q \sim \sqrt{N} \;.
\end{equation}
Finally, key findings about quantum breaking in the PPM continue to hold true in its relativistic generalization \cite{Dvali:2017eba,Kovtun:2020ndc, Kovtun:2020udn,Berezhiani:2020pbv, Berezhiani:2021gph}.

The PPM exhibits two peculiar phenomena. First, a recurrence takes place on short timescales for $\lambda>1$: After quantum breaking has occurred, time evolution periodically returns to a state in which the classical approximation is again viable \cite{Dvali:2013vxa}.\footnote
{Of course, a closed system will generically return close to its initial state but on much longer timescales \cite{Poincare1890,Caratheodry1919}.}
It is reasonable to assume that this behavior is related to a smallness of the Hilbert space. Namely, only three quantum modes are relevant for the dynamics of the PPM if the collective coupling is not much bigger than its critical value, $\lambda \gtrsim 1$ \cite{Dvali:2013vxa}. A second open question arises from the fact that the undercritical regime, $\lambda <1$, does not exhibit quantum breaking, \ie the classical description remains a viable approximation indefinitely \cite{Zell:2019vpe}. In contrast, various arguments indicate that -- at least in certain models -- quantum breaking might also take place far away from a classical instability on the timescale \cite{Dvali:2013eja,Dvali:2017eba,Dvali:2017ruz,Zell:2019vpe} (see also discussion in \cite{Pappalardi:2018frz})

\begin{equation} \label{linearScalingN}
	t_q \sim N \;.
\end{equation}
Since an undercritical choice of parameters can certainly be regarded as more generic, it is important to determine if and under what conditions quantum breaking takes place without a classical instability. 

A comment is in order regarding scrambling. This notion, which was also introduced in the context of black hole physics \cite{Hayden:2007cs,Sekino:2008he}, denotes a generation of entanglement such that information about the initial state can only be extracted by measurements involving a significant part of a quantum system. Clearly, the growth of such quantum correlations indicates a deviation from the classical description and indeed quantum breaking in the PPM can be regarded as a result of scrambling \cite{Dvali:2013vxa} (see also \cite{Hummel:2018brt}).\footnote
{However, these two phenomena are not fully equivalent. For example, in certain situations a breakdown of the classical description may occur due to a loss of entanglement \cite{Almheiri:2012rt,Shenker:2013pqa}. Conversely, quantum breaking can be caused by memory burden instead if entanglement \cite{Dvali:2018xpy, Dvali:2018ytn, Dvali:2020wft}.} 
Studies of scrambling exist in a large variety of other systems; both logarithmic timescales $\sim \ln N$ \cite{Dankert2009,Lashkari:2011yi,Shenker:2013pqa,Maldacena:2015waa,Blake:2016wvh,Maldacena:2016hyu,Shen:2017kez,Aleiner:2016eni,Banerjee:2016ncu,Patel:2016wdy,Bohrdt:2016vhv,Bagrets:2017pwq,Chowdhury:2017jzb, Werman:2017abn,  Rammensee:2018pyk,Chavez-Carlos:2018ijc,Bentsen:2019rlr,Rautenberg:2019kmd,Sunderhauf:2019djv,Li:2020zuj,Belyansky:2020bia,Geiger:2020huu,Kaikov:2022sch,Yin:2022uki} and polynomial scalings $\sim N^\gamma$ \cite{Foss-Feig2014,He2016,Chen2017,Tsuji:2016jbo,Dora:2016gvp,Kukuljan:2017xag,Luitz:2017jrn,Luitz2018,Pappalardi:2018frz,Knap:2018pmj,Chen:2018bqy,Marino:2018hvy,Zhou:2019tde,Yin:2020pjd,Colmenarez:2020juf,Wanisch:2022gyr,Kuwahara:2020chn} have been observed (see \cite{Xu:2022vko,Richter:2022sik} for recent reviews).

Finally, we briefly discuss the classical limit. It is evident from the dependencies \eqref{logarithmicScaling}, \eqref{linearScaling} and \eqref{fractionalScaling} that $\hbar\rightarrow 0$ leads to $t_q\rightarrow\infty$ and no quantum breaking occurs. Since dimensional analysis shows that $N\sim E/(\hbar \omega)$ (see \cite{Dvali:2017eba,Dvali:2017ruz}), where $E$ is the total energy and $\omega$ represents an elementary frequency of (free) oscillations, the classical limit corresponds to an infinite particle number, $N\rightarrow \infty$. Therefore, \eqs \eqref{logarithmicScalingN}, \eqref{rootScalingN} and \eqref{linearScalingN} all  imply $t_q\rightarrow\infty$ and the classical description stays valid indefinitely, as it should.

\paragraph*{Goals \& Outline of Results}
In the present work, our goal is to extend the above studies of quantum breaking in three steps:
\begin{enumerate}
	\item Based on \cite{Kanamoto2002,Dvali:2013vxa}, we shall develop a \textit{new prototype model (NPM)} that features a variable number $Q$ of dynamically accessible modes. In analogy to the PPM, the ground state is determined by a collective coupling $\lambda$. The regime $\lambda>1$ can lead to the logarithmic quantum break-time \eqref{logarithmicScalingN}, but the enlarged Hilbert space of the NPM ensures that recurrence is avoided, \ie the system does not return to the initial classical state on short timescales. Moreover, a large $Q$ causes a breakdown of the classical description in the regime $\lambda<1$, corresponding to the absence of a classical instability, on the linear timescale \eqref{linearScalingN}. As to be discussed, however, such a big number of species puts into question the meaning of quantum breaking \cite{Dvali:2007hz, Dvali:2007wp, Dvali:2008ec}.
	\item We will study how the quantum break-time depends on the various parameters of the NPM. In particular, we shall identify parts of parameter space in which $\lambda>1$ leads to a classical instability but quantum breaking occurs on a polynomial timescale:\footnote
	{A possible sublinear polynomial scaling of the quantum break-time has also been pointed out in \cite{Dvali:2018ytn}.}
\begin{equation} \label{fractionalScalingN}
	t_q \sim N^\gamma \;,
\end{equation}
where numerical fits indicate $\gamma \approx 1/2$.
	\item We will interpolate between $\lambda<1$ corresponding to $t_q \sim N$ and a parameter space in which $\lambda>1$ and $t_q\sim \ln N$. This leads to a continuous transition and a quantum break-time given by \eq \eqref{fractionalScalingN}, where $0<\gamma<1$ depends smoothly on $\lambda$. Close to the critical point $\lambda \approx 1$, we observe $\gamma \approx 1/2$. This corresponds to the timescale \eqref{rootScalingN}, in analogy to the finding in the PPM \cite{Dvali:2015wca}.
\end{enumerate}

The outline of the paper is as follows: After elaborating on the connection of quantum breaking and gravity in section \ref{ssec:analogue}, we will review the PPM of \cite{Kanamoto2002} and its features in section \ref{sec:Periodic}. Subsequently, we present in section \ref{sec:Model} the NPM and derive analytic estimates related to the validity of the classical description. In section \ref{sec:numeric}, we numerically compute real-time evolution to determine how the quantum break-time of the NPM depends on its different parameters, where we use the computer program \textit{TimeEvolver} \cite{Michel:2022kir}.\footnote
{\textit{TimeEvolver} relies on Krylov-subspace methods for computing real-time dynamics. Similar techniques have \eg recently also been employed in \cite{Khlebnikov:2013yia,Varma2015,Luitz2015,EkinKocabas2016,Rehn2016,Luitz:2017jrn,Luitz2018,Knap:2018pmj,Colmenarez:2020juf,Lang:2022wtd,Halimeh:2022rwu} (see discussion in \cite{Luitz2016}).}
In particular, we investigate how our system interpolates between the undercritical case $\lambda <1$ and the overcritical regime of $\lambda >1$. Finally, we conclude in section \ref{sec:Conclusion} and give an outlook to implications of our results for gravity.

\subsection{Analogue models for gravity}
\label{ssec:analogue}
As discussed above, an important motivation for studying quantum breaking originates from open questions related to black holes. Therefore, we shall give a brief review of selected issues in high-energy physics.  Classically, black holes are stable and (in the absence of rotation) static, but once quantum effects are included they emit Hawking radiation \cite{Hawking:1975vcx}. The computation of \cite{Hawking:1975vcx} is semi-classical, \ie quantum fields are studied in the background of a classical and fixed black hole metric. Nevertheless, it is clear that the produced particles must backreact on the black hole, not least because of energy conservation. 
However, how exactly the black hole evolves as a result of Hawking evaporation is an open question  \cite{Hawking:1976ra}. As one option, it is possible that its classical description breaks down once a sufficient number of particles has been emitted \cite{Page:1979}. In particular, it appears that unitarity is violated if black hole evaporation can still be computed semi-classically after half of its mass is lost \cite{Page:1993wv}.

The purpose of the quantum N-portrait \cite{Dvali:2011aa} is to address these issues in a microscopic model. In it a black hole of mass $M$ and Schwarzschild radius $r_g = 2 G_N M$ consists of $N= G_N M^2/\hbar$ soft gravitons, where $G_N$ is Newton's constant. These constituent gravitons have wavelength $r_g$ and interact with coupling strength $\alpha = \hbar G_N/r_g^2 = 1/N$. Therefore, the collective coupling gives $\lambda = \alpha N = 1$ and quantum criticality appears as a crucial property of black holes \cite{Dvali:2011aa,Dvali:2012rt,Dvali:2012en,Dvali:2012wq,Dvali:2013vxa}. Studying the PPM close to the critical point $\lambda=1$ yielded indications that the classical description of a black hole ceases to be valid at the latest after half of the mass is lost \cite{Dvali:2012en,Dvali:2015ywa, Dvali:2015wca, Dvali:2016zqx}. In particular, such a finding ensures compatibility with unitary time evolution. 

The PPM and the approach of \cite{Dvali:2012en} led to the development of new non-relativistic analogue systems that share important properties with black holes \cite{Dvali:2018tqi, Dvali:2018xpy, Dvali:2020wft}. Whereas previous proposals focused on imitating geometry \cite{Unruh:1980cg,Unruh:1994je, Visser:1997ux, Garay:2000jj, Barcelo:2001ca} (see \cite{Barcelo:2005fc,Jacquet:2020bar} for reviews), these new models reproduce information-theoretic characteristics of a black hole, in particular its Bekenstein-Hawking entropy \cite{Bekenstein:1973ur}. Such an approach is especially useful for gathering information about the fate of a black hole after its half lifetime: Hints were found that black hole evaporation can slow down drastically once its classical description has broken down \cite{Dvali:2020wft}.

Related issues also arise in expanding spacetimes, especially in the de Sitter solution of General Relativity. Classically, this system is eternal but quantum effects also cause particle production \cite{Gibbons:1977mu}, in close analogy to Hawking radiation. Again this leads to a question about how the generated particles backreact on the de Sitter spacetime \cite{Mottola:1984ar, Ford:1984hs, Mottola:1985, Antoniadis:1985pj}. As in the black hole case, this issue is also linked to unitarity \cite{Danielsson:2002td,Goheer:2002vf,Banks:2003pt, Arkani-Hamed:2007ryv}. Moreover, there are proposals for employing analogue quantum systems for studying quantum effects in an expanding Universe \cite{Fedichev:2003bv, Fedichev:2003id, Barcelo:2003et, Weinfurtner:2004mu, Fischer:2004bf, Uhlmann:2005hf}, where as in \cite{Unruh:1980cg} the focus is on imitating geometric properties.

 Similarly to the quantum N-portrait \cite{Dvali:2011aa} of a black hole, one can develop a microscopic model of de Sitter and inflationary spacetimes \cite{Dvali:2013eja,Berezhiani:2016grw,Dvali:2017eba,Dvali:2020etd,Berezhiani:2021zst}:\footnote
 {See also \cite{Dvali:2022vzz} for an analogous approach for resolving classical backgrounds in QED.}
  Now the constituent gravitons have wavelength $H^{-1}$ and their number is $N=1/(\hbar G_N H^2)$, where $H$ is the Hubble rate. Since the strength of gravitational interactions is given by $\alpha=\hbar G_N H^2$, again $\lambda=1$ and criticality emerges as a crucial property. Moreover, the microscopic model strongly deviates from the description in terms of a classical metric once the backreaction of produced particles becomes sizable \cite{Dvali:2013eja,Dvali:2017eba}. These studies were developed further to include information-theoretic aspects, in particular the Gibbons-Hawking entropy \cite{Gibbons:1977mu} of de Sitter. As a result, hints emerged for the existence of new observables for inflation that are invisible in the semi-classical description and that are sensitive to the whole duration of inflationary expansion \cite{Dvali:2018ytn}.
 	
	\section{Review of attractive one-dimensional Bose gas}\label{sec:Periodic}
\subsection{Full model}\label{sec:PeriodicModel}

Following the analysis of \cite{Kanamoto2002}, we shall review the model of an attractively interacting cold Bose gas on a 1-$d$ circle. Its Hamiltonian is given by:\footnote
{We choose $\hat{\psi}$ to be dimensionless such that $\left[\hat{\psi}(\theta), \hat{\psi}^\dagger(\theta')\right]=\delta(\theta-\theta')$.}
\begin{equation}\label{periodicHamiltonianPos}
	\hat{H} = \frac{\hbar^2}{2 m R^2} \int_0^{2\pi} \diff\theta \left[-\hat{\psi}^\dagger(\theta) \partial_{\theta}^2 \hat{\psi}(\theta) - \frac{\pi \alpha}{2} \hat{\psi}^\dagger (\theta) \hat{\psi}^\dagger (\theta)  \hat{\psi} (\theta) \hat{\psi} (\theta) \right] \, , 
\end{equation}
where $R$ is the radius of the circle, $\alpha$ a dimensionless coupling constant and $m$ corresponds to the particle mass.
Throughout this paper we will assume $\alpha > 0$. From here on, we shall choose units of energy such that $\hbar^2/(2 m R^2)=1$.
In the mean field approximation, we split the (time-dependent) wave function in a classical contribution and fluctuations, $\hat{\psi}(t,\theta) = \psi_0(t,\theta) + \hat{\delta \psi}(t,\theta)$. The solution of the mean field is determined by the Gross-Pitaevskii equation: 
	\begin{equation} 	\label{GP}
		i \hbar \partial_t \psi_0(t,\theta)    = 
		\left( - \partial_{\theta}^2    -    \pi \alpha  |\psi_0(t,\theta)|^2
		\right) \psi_0(t,\theta) = \mu \psi_0(t,\theta) \,, 
\end{equation}
where $\mu$ is a Lagrange multiplier enforcing the constraint of particle number conservation, $\int_0^{2\pi} d\theta |\psi_0(t,\theta)|^2 = N$.

For periodic boundary conditions we can obtain two stationary solutions, $\psi_0(t,\theta)=\text{e}^{-i\mu t/\hbar}\psi_0(\theta)$, from \eq \eqref{GP} \cite{Carr2000, Kanamoto2002}. The first one is a homogeneous condensate,
	\begin{equation}\label{groundstateGP1}
	\psi^{(<)}_0(\theta) = 
		\sqrt{\frac{N}{2 \pi}}  \;.
\end{equation}
The second one only exists for $\lambda>1$ and corresponds to a bright soliton,
\begin{equation}\label{groundstateGP2}
	\psi^{(>)}_0(\theta) = 
\sqrt{\frac{N K(m)}{2\pi E(m)}} dn \left( \frac{K(m)}{\pi} (\theta -\theta_0) | m\right) \;,
\end{equation}
where $dn(u|m)$ is a Jacobian elliptic function and $K(m)$ as well as $E(m)$ correspond to the complete elliptic integrals of the first and the second kind, respectively. Finally, $m$ is determined by the equation
\begin{equation}\label{determineM}
	K(m) E(m) = (\pi/2)^2 \lambda \;,
\end{equation} 
and we introduced the collective coupling
\begin{equation} \label{collectiveCoupling}
	\lambda = \alpha N \;.
\end{equation} 
For small $\lambda < 1$, the ground state is given by \eq \eqref{groundstateGP1}. For large $\lambda > 1$, the homogeneous condensate \eqref{groundstateGP1} becomes unstable and the energy functional is minimized by the solution \eqref{groundstateGP2} with center $\theta_0$. Both phases are separated by a second order phase transition at $\lambda=1$.

Rather than working in position space, it will be more convenient to go to momentum space. To this end, we expand the field operators $\hat{\psi}$ in Fourier modes
\begin{equation} \label{modeExpansion}
	\hat{\psi}(\theta) = \frac{1}{\sqrt{2 \pi}} \sum_{k = - \infty}^\infty \hat{a}_k \text{e}^{i k \theta},
\end{equation} 
where the creation and annihilation operators $\hat{a}_k^{\dagger}$ of (angular) momentum mode $k$ satisfy the usual canonical commutation relations
\begin{equation}
	[\hat{a}_j,\hat{a}_k^\dagger] = \delta_{kj} \, , \hspace{20pt} [\hat{a}_j,\hat{a}_k] = [\hat{a}_j^\dagger, \hat{a}_k^\dagger ] = 0 \, . 
\end{equation}
By plugging this expansion into the Hamiltonian \eqref{periodicHamiltonianPos}, we arrive at: 
\begin{equation} \label{periodicFull}
	\hat{H} = \sum_{k=-\infty}^\infty k^2 \hat{n}_{k} - \frac{\alpha}{4}\sum_{k,l,m = -\infty}^\infty \hat{a}_k^\dagger \hat{a}_l^\dagger \hat{a}_{m+k} \hat{a}_{l-m} \;.
\end{equation}

In the present paper, a homogeneous condensate, \ie a state in which only the $\hat{a}_0$-mode is occupied, 	
\begin{equation} \label{initialState}
	\ket{\text{in}} = \ket{N, 0, \ldots, 0} \;,
\end{equation}
plays a special role and will be used a initial state for time evolution.. Throughout we will assume $N\gg 1$ so that we can employ the Bogoliubov approximation, $\hat{a}_0 \approx \sqrt{N}$, to get \cite{Dvali:2012en,Flassig:2012re}:
\begin{equation} \label{periodicFullBogo}
	\hat{H} =  \sum_{k\neq 0} \left(k^2 - \frac{\lambda}{2}\right) \hat{n}_{k} -\frac{\lambda}{4} \sum_{k\neq 0} \left(\hat{a}_k^\dagger \hat{a}_{-k}^\dagger + \hat{a}_k \hat{a}_{-k} \right) \;.
\end{equation} 
In \eq \eqref{periodicFullBogo}, we neglected contributions of order $\hat{a}_k^4$ for $k\neq 0$. This leads to an error that scales like $1/N$.

The reason for the choice \eqref{initialState} of initial state is twofold. First, $\ket{\text{in}}$ corresponds to the mean-field solution \eqref{groundstateGP1}, which is energetically favorable for $\lambda<1$. Even though $\ket{\text{in}}$ has no connection to the ground state for $\lambda>1$, its classical limit, corresponding to the homogeneous condensate \eqref{groundstateGP1}, is still a mean field solution, albeit an unstable one. Therefore, it does not evolve in time, as long as no symmetry-breaking perturbations are included.\footnote
{The formation of the inhomogeneous soliton is discussed in \cite{Kanamoto2004,Kanamoto2005, kanamoto2009, kanamoto2010}.}
Consequently, any time evolution away from the initial state inevitably leads to a deviation from classicality for all values of $\lambda$, as discussed in \cite{Dvali:2013vxa,Hummel:2018brt,Rautenberg:2019kmd}.
The second reason for choosing $\ket{\text{in}}$ is that one can make an inference from the occupation number of the $\hat{a}_0$-mode to the full quantum state: A state is close to \eq \eqref{initialState} if and only if the occupation number of the $\hat{a}_0$ is close to $N$. As a result, evaluating the expectation value of $\hat{n}_0$ suffices for identifying quantum breaking.

\subsection{Undercritical regime}\label{sec:PeriodicUnder}
For diagonalizing the system, we perform the Bogoliubov transformation \cite{Dvali:2012en,Flassig:2012re}
\begin{equation} \label{bogoTrafoDefinition}
	\hat{a}_k = u_k \hat{b}_k + v_k^*\hat{b}_{-k}^\dagger \,,
\end{equation}
where we set
\begin{equation} \label{bogoTrafoPeriodic}
	u_k^2 = \frac{1}{2}\left(\frac{k^2-\frac{\lambda}{2}}{\epsilon_k}+1\right) \;, \qquad
	v_k^2 = \frac{1}{2} \left(\frac{k^2-\frac{\lambda}{2}}{\epsilon_k} -1\right)\;,
\end{equation}
and defined
\begin{eqnarray} \label{energyBogoliubov}
	\epsilon_k = \sqrt{k^2\left(k^2 - \lambda\right)} \;.
\end{eqnarray}
This yields
\begin{equation} \label{periodicBogoliubovFull}
	\hat{H} = \sum_{k\neq 0} \epsilon_k \hat{n}^b_k \;.
\end{equation}
Evidently, the Bogoliubov transformation defined by \eq \eqref{bogoTrafoPeriodic} only exists in the undercritical case $\lambda < 1$. We shall first analyze the system in this parameter regime. 

Our goal is to decide whether quantum breaking can take place in this case. To this end, we need to determine by how much time evolution can take the system away from the initial state \eqref{initialState}. We can obtain an estimate by considering the ground state $\ket{0_b}$ of the Bogoliubov modes. Computing the number of depleted particles, \ie particles that are not in the $\hat{a}_0$-mode, yields
\begin{equation} \label{depletedFormula}
	N_d = \sum_{k\neq 0} \bra{0_b} \hat{n}_k \ket{0_b} =  \sum_{k\neq 0} |v_k|^2 = \frac{1}{2} \sum_{k\neq 0} \left(\frac{k^2-\frac{\lambda}{2}}{\sqrt{k^2\left(k^2 - \lambda\right)}} -1\right) \;.
\end{equation}
For $\lambda \ll 1$, we obtain \cite{Zell:2019vpe}
\begin{equation} \label{depletedUndercritical}
	N_d  = \frac{\pi^4 \lambda^2}{720} \approx 0.135 \lambda^2 \;.
\end{equation} 
We conclude that depletion is small, $N_d \ll 1 \ll N$, in the undercritical regime. We must emphasize, however, that time evolution in a closed system will never lead to the Bogoliubov ground state. Therefore, the number \eqref{depletedUndercritical} can only be viewed as a qualitative estimate of the possible magnitude of deviations from the initial state \eqref{initialState}. Nevertheless, \eq \eqref{depletedUndercritical} clearly indicates that starting from the initial state \eqref{initialState}, most particles will stay in the zero mode. Thus, time evolution will not lead to a large departure from the initial state and quantum breaking cannot take place. As a second inference from \eq \eqref{depletedUndercritical}, we observe that only the modes $k=\pm1$ contribute significantly to depletion. If we only consider these two modes, we get
\begin{equation} \label{depletedUndercriticalThreeModes}
	N_d^{(1)}  = \frac{\lambda^2}{8} =0.125 \lambda^2 \;,
\end{equation} 
which is very close to the total number $N_d$ of depleted particles.

\begin{figure}
	\centering 
	\includegraphics[width=0.4\linewidth]{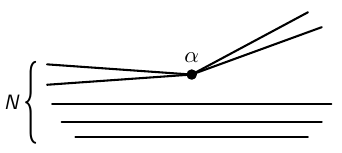}
	\caption{Schematic depiction of the leading-order process by which an initial state (on the left), in which are particles are in the $\hat{a}_0$-mode, evolves towards a state (on the right), in which two particles have left the $\hat{a}_0$-mode.}
	\label{fig:scattering}
\end{figure}

Even though no significant departure from the initial state \eqref{initialState} will occur, we can estimate the speed at which particles leave the zero modes in the very first moments of time evolution. To this end, we consider a process in which two particles of the $\hat{a}_0$-modes annihilate and produce two particles in a mode with non-zero $k$ (see \fig \ref{fig:scattering}). As is evident from \eq \eqref{periodicFull}, the amplitude of such a process scales as $\alpha$. This leads to the rate (see also \cite{Dvali:2012rt,Dvali:2012en,Dvali:2012wq,Dvali:2013eja,Dvali:2017eba})
\begin{equation} \label{rateUndercritical}
	\Gamma = N^2 \alpha^2 = \lambda^2 \;,
\end{equation}
where the enhancement of $N^2$ is due to the fact that in the state \eqref{initialState}, there are approximately $N^2$ possibilities for choosing the two particles of the $\hat{a}_0$-mode that annihilate. Generically, the rate \eqref{rateUndercritical} receives additional contributions with higher powers of $\lambda$. Thus, this formula is only valid if $\lambda$ is sufficiently smaller than $1$. If the process of generating particles with non-zero $k$ were to continue -- which it does not -- with the rate \eqref{rateUndercritical}, then we would be able to estimate the resulting quantum break-time. Namely, the number $\Delta N$ of particles that have departed the $\hat{a}_0$-mode would be given as
\begin{eqnarray}
	\Delta N = \Gamma t \;.
\end{eqnarray}
This would result in the quantum break-time as the timescale when $\Delta N \sim N$:
\begin{equation} \label{breaktTimeUndercritical}
	t_{\text{q}} = N \Gamma^{-1} = \frac{N}{\lambda^2} \;.
\end{equation}
We must reiterate, however, that the timescale \eqref{breaktTimeUndercritical} is fictitious since quantum breaking does not take place in the system \eqref{periodicFull} for $\lambda < 1$.

\subsection{Overcritical regime}\label{sec:PeriodicOver}
The Bogoliubov transformation employed in the previous section is only valid in the undercritical regime $\lambda < 1$. In order to study excitations in the overcritical case, we shall include perturbations in the Gross-Pitaevskii equation \eqref{GP}. To leading order in $\hat{\delta \psi}(t,\theta)$, we get \cite{Kanamoto2002}: 
\begin{align}
	\left(- \partial_{\theta}^2    -   2 \pi \alpha   |\psi_0(t,\theta)|^2 - \mu\right) \hat{\delta \psi}(t,\theta) -  \pi \alpha  \psi_0^2(t,\theta) \hat{\delta \psi}^\dagger(t,\theta)  & = \omega_k \hat{\delta \psi}(t,\theta) \;,\\
	\left(- \partial_{\theta}^2    -   2 \pi \alpha  |\psi_0(t,\theta)|^2 - \mu\right) \hat{\delta \psi}^\dagger(t,\theta) -  \pi \alpha \psi_0^{\dagger 2}(t,\theta) \hat{\delta \psi}(t,\theta)  & = - \omega_k \hat{\delta \psi}^\dagger(t,\theta) \;,
\end{align}
where $\omega_k$ are the Bogoliubov-de Gennes frequencies \cite{Bogolyubov:1947zz, deGennes1999}. For the homogeneous solution \eqref{groundstateGP1}, we have $|\psi_0|^2 = \sqrt{N}/\sqrt{2\pi}$ and correspondingly $\mu = - \alpha N/2$. Using the mode expansion \eqref{modeExpansion} for $k\neq 0$, we arrive at \cite{Kanamoto2002} (see also \cite{Dvali:2017ruz})
\begin{equation} \label{BdGFrequencies}
	\omega_k = \sqrt{k^2(k^2-\lambda)} \;.
\end{equation}
For $\lambda <1$, this agrees with the result \eqref{energyBogoliubov} of the Bogoliubov transformation.
However, the result \eqref{BdGFrequencies} is also valid in the overcritical regime, \ie for $\lambda > 1$. In this case, the lowest-lying frequency becomes imaginary,
\begin{equation} \label{imaginaryFrequency}
	\omega_1 = i \sqrt{\lambda-1} \;,
\end{equation}
which signals the presence of an instability.

From this finding we can derive a rough estimate of the quantum break-time. In order to account for the instability, a contribution should appear in the depletion rate $\Gamma$ that scales with the number $\Delta N$ of particles that have already left the $\hat{a}_0$-mode. This effects should also depend on the magnitude of the imaginary frequency \eqref{imaginaryFrequency} and vanish for $\omega_1\rightarrow 0$. Correspondingly, we modify the rate \eqref{rateUndercritical} as:
\begin{equation} \label{rateOvercritical}
	\Gamma = \lambda^2 + \sqrt{\lambda - 1} \Delta N\;,
\end{equation}
where we employed a linear power of $\omega_1$, as also in \cite{Dvali:2013vxa}. Needless to say, this formula only represents a rough estimate and its implications remain to be verified. In particular, the derivation of the first term $\sim \lambda^2$ is only valid for $\lambda<1$, but we expect that extrapolating is possible since the rate of perturbative scattering does not change qualitatively for $\lambda \gtrsim 1$. In any case, we are interested in the case when the second term in \eq \eqref{rateOvercritical} dominates. Then the number of depleted particles becomes
\begin{equation}
	\Delta N = \text{e}^{\sqrt{\lambda - 1} t} \;.
\end{equation}
Consequently, $\Delta N \sim N$ leads to fast quantum breaking after the timescale  \cite{Dvali:2013vxa} (\cf \eq \eqref{logarithmicScalingN})
\begin{equation} \label{breaktimeOvercritical}
	t_q = \frac{\ln N}{\sqrt{\lambda - 1}} \;.
\end{equation}

As in the undercritical case, we will finally study the number $N_d$ of particles that are outside the $\hat{a}_0$-mode in the ground state. Since for $\lambda >1$ the homogeneous state \eqref{initialState} is no longer related to the ground state \eqref{groundstateGP2}, we expect this number to be macroscopic. Indeed, we can derive from \eq \eqref{modeExpansion} in the mean-field approximation:
	\begin{equation}
		a_k = \frac{1}{\sqrt{2\pi}} \int \diff \theta\, \text{e}^{-i k \theta} \psi_0(\theta) \;.
	\end{equation}
Whereas the homogeneous condensate \eqref{groundstateGP1} would lead to $a_k \sim \delta_{k0} \sqrt{N}$, the ground state for $\lambda>1$ is given by the bright soliton \eqref{groundstateGP2}. Thus, $a_k \sim \sqrt{N}$ at least for one $k \neq 0$ and we conclude that
\begin{equation} \label{depletionOvercritical}
	N_d \sim N \;.
\end{equation}
	Even though the ground state can never be reached dynamically in a closed system, this indicates that quantum breaking generically takes place for $\lambda>1$, in agreement with previous findings \cite{Dvali:2013vxa}. Starting from the initial state \eqref{initialState}, we can estimate how many particles leave the $\hat{a}_0$-mode. Namely, the instability disappears due to backreaction, \ie that fact that particles leaving the $\hat{a}_0$-mode effectively decrease the collective coupling $\lambda$ (see also \cite{Kanamoto2002}). This yields
	\begin{equation} \label{turningPointOvercritical}
		\sqrt{\lambda\left(1-\frac{\Delta N}{N}\right)}=1 \;, \qquad \Rightarrow \qquad \Delta N = N \frac{\lambda - 1}{\lambda} \;.
	\end{equation}

\subsection{Critical regime}
At the critical point $\lambda=1$, we can obtain an estimate for the number of depleted particles by extrapolating the result \eqref{depletedFormula} of the undercritical regime. To this end, we take into account backreaction to derive from \eq \eqref{depletedFormula}:
	\begin{equation} \label{depletedPeriodicBackreaction}
		N_d =  \frac{1-\frac{\lambda}{2}\left(1-\frac{N_d}{N}\right)}{\sqrt{1-\lambda\left(1-\frac{N_d}{N}\right)}} -1 \;,
	\end{equation}
	where we neglected all modes $|k|>1$. This is possible since they are still undercritical at $\lambda=1$ and so their contribution to depletion can still be estimated by \eq \eqref{depletedUndercritical}. 
	For $N_d\ll N$, \eq \eqref{depletedPeriodicBackreaction} implies \cite{Kanamoto2002} (see also \cite {Panchenko:2015dca}):
	\begin{equation} \label{depletionCritical}
		N_d = \frac{N^{1/3}}{2^{2/3}} \left(1+O\left(\frac{1}{N^{2/9}}\right)\right) \;.
	\end{equation}
Analytic arguments relying on the limit of large $N$ as well as numerical analysis indicate that the quantum break-time scales as \cite{Dvali:2015wca}
\begin{equation} \label{breaktimeCritical}
	t_q \sim \sqrt{N} \;.
\end{equation}

\subsection{Truncation}
As is evident from the previous discussion, depletion is mostly due to the modes $k=\pm1$ for all values of $\lambda$ (as long as $\lambda < 4$). Therefore, we shall consider the truncation of the model \eqref{periodicFull} to three modes:
\begin{equation} \label{periodicThreeModes}
	\hat{H} = \hat{n}_{-1} + \hat{n}_1 - \frac{\alpha}{4}\sum_{k,l = -1}^1\sum_{m=\max(-k-1,l-1)}^{\min(-k+1,l+1)} \hat{a}_k^\dagger \hat{a}_l^\dagger \hat{a}_{m+k} \hat{a}_{l-m} \;.
\end{equation}
The motivation to consider this system is simplicity or equivalently better suitability for numerical study. Since modes with $k\geq 2$ do not contribute significantly to quantum breaking, we can discard them from the outset.

In the case $\lambda < 1$, the truncation of Hamiltonian \eqref{periodicFullBogo} after Bogoliubov approximation leads to
\begin{equation} \label{periodicThreeModesBogo}
	\hat{H} =  \left(1 - \frac{\lambda}{2}\right) \left(\hat{n}_{-1} + \hat{n}_{1}\right)  -\frac{\lambda}{2}  \left(\hat{a}_1^\dagger \hat{a}_{-1}^\dagger + \hat{a}_1 \hat{a}_{-1} \right) \;,
\end{equation} 
and the Bogoliubov Hamiltonian \eqref{periodicBogoliubovFull} becomes
\begin{equation} \label{periodicBogoliubovThreeModes}
	\hat{H} =  \sqrt{1 - \lambda} \left(\hat{n}^b_{-1} +  \hat{n}^b_1\right) \;.
\end{equation}
For $\lambda \lesssim 1$, the number of depleted particles is still given by \eq \eqref{depletedPeriodicBackreaction}, where we took into account backreaction.
	For $\lambda < 1$, the solution is
	\begin{equation} \label{depletionPeriodic}
		N_d =	\left(\frac{1-\frac{\lambda }{2}}{\sqrt{1-\lambda }}-1\right) \left(1-\frac{\lambda ^2}{4 (1-\lambda )^{3/2} N} + O\left(\frac{1}{N^2}\right)\right) \;. 
	\end{equation}
	In the critical and overcritical cases, the results \eqref{depletionCritical} and \eqref{depletionOvercritical} remain valid.

\section{New prototype model}\label{sec:Model}
\subsection{Extension by species}
Now we shall consider a modified version of the model \eqref{periodicThreeModes}:\footnote{A related simplification was considered in \cite{Dvali:2017nis,Dvali:2018vvx,Dvali:2018tqi}.} 
\begin{equation} \label{simplified}
	\hat{H} =  \hat{n}_1 - \frac{\alpha}{4} \left(2 \hat{n}_0\hat{n}_1 + \hat{a}_0^{\dagger\,2} \hat{a}_1^2  +  \hat{a}_1^{\dagger\,2} \hat{a}_0^2\right) \;.
\end{equation}
Again using the state \eqref{initialState}, we get in the Bogoliubov approximation, $\hat{a}_0 \approx \sqrt{N}$,
\begin{equation} \label{simplifiedBogo}
	\hat{H} =  \left(1-\frac{\lambda}{2}\right)\hat{n}_1 - \frac{\lambda}{4} \left(\hat{a}_1^2  +  \hat{a}_1^{\dagger\,2} \right) \;,
\end{equation}
where we used the collective coupling $\lambda$ defined in \eq \eqref{collectiveCoupling} and again neglected an error that scales as $1/N$. We already observe that \eq \eqref{simplifiedBogo} resembles its counterpart in the truncation of the periodic system, as shown in \eq \eqref{periodicThreeModesBogo}. For $\lambda <1$, we can employ a Bogoliubov transformation analogous to \eq \eqref{bogoTrafoPeriodic}:
\begin{equation} \label{bogoTrafo}
	u^2 = \frac{1}{2}\left(\frac{1-\frac{\lambda}{2}}{\sqrt{1 - \lambda}}+1\right) \;, \qquad
	v^2 = \frac{1}{2} \left(\frac{1-\frac{\lambda}{2}}{\sqrt{1 - \lambda} } -1\right)\;.
\end{equation}
We arrive at
\begin{equation} \label{simplifiedBogoliubovFull}
	\hat{H} = \sqrt{1 - \lambda} \, \hat{n}^b \;,
\end{equation}
which coincides with the Bogoliubov Hamiltonian of either of the $|k|=1$-modes of the truncation of the periodic system to three modes (see \eq \eqref{periodicBogoliubovThreeModes}).This represents a motivation for constructing our prototype model as shown in \eq \eqref{simplified}.

For $\lambda < 1$, it follows from \eq \eqref{simplifiedBogo} that depletion is given by 
\begin{equation} \label{depletionSimplifiedOneMode}
	N_d =  \frac{1}{2}\left(\frac{1-\frac{\lambda}{2}\left(1-\frac{N_d}{N}\right)}{\sqrt{1-\lambda\left(1-\frac{N_d}{N}\right)}} -1\right) \;,
\end{equation}
where as above we took into account backreaction.
This coincides with the result \eqref{depletedPeriodicBackreaction} of the truncated periodic system, up to a factor of $1/2$, which accounts for the fact that the system \eqref{simplified} only has one non-zero mode instead of two. Therefore, depletion is still small and quantum breaking does not take place in the undercritical regime of the system \eqref{simplified}.

In order to enhance depletion, we shall introduce $Q$ species for the $\hat{a}_1$-mode:\footnote
{This system is very similar to the full periodic model \eqref{periodicFull}, if one replaces $k^2 \rightarrow 1$ in the kinetic terms. So one might also considering using such a periodic model with modified energy gaps.}

\begin{equation} \label{simplifiedSpecies}
	\boxed{\begin{aligned} 
			\hat{H}& =  \sum_{k=1}^Q\left(\hat{n}_k - \frac{\alpha}{4} \left(2 \hat{n}_0\hat{n}_k + \hat{a}_0^{\dagger\,2} \hat{a}_k^2  +  \hat{a}_k^{\dagger\,2} \hat{a}_0^2\right)\right)\\
			&+\frac{C_m}{2}\sum_{k=1}^{Q}\sum_{\substack{l=k+1}}^{Q} f(k,l)\left(\hat{a}_{k}^{\dagger\, 2} \hat{a}_{l}^2 +  \text{h.c.}\right) \;.
	\end{aligned}}
\end{equation}
This Hamiltonian defines our new prototype model (NPM) that we shall study in the following. As is evident from \eqs \eqref{simplifiedBogo} and \eqref{simplifiedBogoliubovFull}, exclusively occupying the $\hat{a}_0$-mode has the same effect in our new prototype model \eqref{simplifiedSpecies} as in the periodic system \eqref{periodicFull}: For $\lambda<1$, the corresponding state \eqref{initialState} coincides with the ground state in the mean-field limit where a classical instability appears for $\lambda >1$.

A final remark is in order concerning the second line of \eq \eqref{simplifiedSpecies}. In an analogous approach as in \cite{Dvali:2020wft}, we introduced this interaction to ensure that the system possesses no symmetries other than the conservation of total particle number. We choose the same $f(k,l)$ as in \cite{Dvali:2020wft}, which takes essentially random values in $|f(k, l)|\in [0.5;1]$ with both positive and negative signs.\footnote
{Concretely, the function is \cite{Dvali:2020wft}: 
	$f(k,l) = 
	\left\lbrace \begin{array}{rcl}
		F(k,l) -1 & \mbox{for} & F < 0.5
		\\ 
		F(k,l) & \mbox{for} & F\geq 0.5 
	\end{array}\right.$, where $F(k,l) = \left( \sqrt{2}(k-1)^3 + \sqrt{7}(l-1)^5 \right) \mod 1$. We remark, however, that the structure of the self-interaction of the $\hat{a}_k$-modes is different as compared to \cite{Dvali:2020wft}.}
As we shall demonstrate subsequently, the value of $C_m$ does not influence the system in any essential way, provided that $C_m$ is not too large. We give some analytic estimates regarding this requirement in appendix \ref{app:Cm}. In the numerical study, we shall moreover check empirically that the values of $C_m$ that we consider do not change the qualitative behavior of the system.

\subsection{Undercritical regime}\label{sec:ModelUnder}
As long as $C_m$ is not too large, we can approximate the $\hat{a}_k$-modes, $k\geq 1$, as decoupled from each other. Then for each mode, the Bogoliubov approximation is still given by \eq \eqref{simplifiedBogo} so that the number of depleted particles is enhanced by a factor of $Q$ as compared to \eq \eqref{depletionSimplifiedOneMode}:
\begin{equation} \label{depletionSimplified}
	N_d = \frac{Q}{2} \left(\frac{1-\frac{\lambda}{2}\left(1-\frac{N_d}{N}\right)}{\sqrt{1-\lambda\left(1-\frac{N_d}{N}\right)}} -1\right) \;.
\end{equation}
For $N_d\ll N$, the solution is
\begin{equation} \label{depletion}
	N_d =	\frac{Q}{2} \left(\frac{1-\frac{\lambda }{2}}{\sqrt{1-\lambda }}-1\right) \left(1-\frac{\lambda ^2 Q}{8 (1-\lambda )^{3/2} N} + O\left(\frac{N_d^2}{N^2}\right)\right) \;. 
\end{equation}
This clearly indicates that quantum breaking can take place for sufficiently large $Q$.\footnote
{Additionally, we can read off from \eq \eqref{depletion} an estimate for the coefficient of corrections that scale as $1/N$. We get $\lambda^2 Q / (8 (1-\lambda )^{3/2})$.}
For $\lambda \ll 1$, we can approximate
\begin{equation} \label{depletionApproximate}
	N_d \approx \frac{Q \lambda^2}{16} \;.
\end{equation}
in accordance with \eq \eqref{depletedUndercriticalThreeModes}.

We proceed to make an estimate of the quantum break-time. As in the periodic system, we consider the process in which two particles of the $\hat{a}_0$-mode annihilate to produce two excitations with $k \geq 1$. The introduction of $Q$ channels in the final state enhances the rate \eqref{rateUndercritical} as
\begin{equation} \label{rateUndercriticalSpecies}
	\Gamma = Q N^2 \alpha^2 = Q \lambda^2 \;. 
\end{equation}
Consequently, the timescale after which a significant numbers of particles has left the $\hat{a}_0$-mode is (see also \cite{Dvali:2017eba,Dvali:2020etd,Dvali:2021bsy} for the effect of species on the quantum break-time)
\begin{equation} \label{breaktimeUndercriticalSpecies}
	t_q = N \Gamma^{-1} = \frac{N}{Q  \lambda^2} \;,
\end{equation}
in accordance with \eq \eqref{breaktTimeUndercritical}.
 At first sight, it appears that we have constructed a system that exhibits quantum breaking without any classical instability.

Comparing \eqs \eqref{depletionApproximate} and \eqref{breaktimeUndercriticalSpecies} reveals an immediate problem.  Quantum breaking corresponding to a full deviation from the classical solution requires $N_d \sim N$, which leads to $Q \gtrsim N/\lambda^2$. Once $Q$ is so large, however, \eq \eqref{breaktimeUndercriticalSpecies} implies $t_q \lesssim 1$. This means that the quantum break-time is shorter than the timescale of a single free oscillation. (We recall that our choice of units is such that $\hbar^2/(2 m R^2)=1$.) Therefore, one cannot say that a viable classical description existed to begin with and the meaning of quantum breaking becomes unclear. This issue is particularly evident in the classical limit $\hbar\rightarrow 0$. Generically, we know the scaling $N\sim \hbar^{-1}$ (see \cite{Dvali:2017eba,Dvali:2017ruz}). Since full quantum breaking requires $Q \gtrsim N/\lambda^2$, it follows that $Q\sim N \sim \hbar^{-1}$. Then \eq \eqref{breaktimeUndercriticalSpecies} implies that $t_q \sim \hbar^0$, \ie the quantum break-time would not vanish in the classical limit $\hbar\rightarrow 0$. Clearly, this does not make sense.

The above argument was first applied in quantum gravity \cite{Dvali:2007hz, Dvali:2007wp, Dvali:2008ec}: Requiring that at least one emission of a Hawking quantum can be well-described semi-classically leads to an upper bound on the number of hidden species. Recently, this finding was connected to a breakdown of perturbative unitarity \cite{Dvali:2019jjw, Dvali:2019ulr,Dvali:2020wqi} (see \cite{Dvali:2021rlf,Dvali:2021tez} for concrete models). Namely, taking into account $Q$ species effectively enhances coupling constants by powers of $Q$. Therefore, perturbation theory only remains valid if $Q$ is not too large.\footnote
{Applied to our system, it appears that including $Q$ species has the same effect as replacing the coupling $\alpha \rightarrow \alpha_{\text{eff}} = \alpha \sqrt{Q}$ (see \eg the process shown in \fig \ref{fig:scattering}). Then $Q \sim N/\lambda^2$ leads to $\alpha_{\text{eff}} = 1/\sqrt{N}$, which still allows for a perturbative expansion.}

For our present discussion, the availability of a perturbative approximation is not essentially since we shall study our system by exact non-perturbative means. As discussed above, however, $t_q \ll 1$ would put into question the meaning of quantum breaking. Still, there are several caveats in the above argument. First, it is unclear if in all parts of parameter space the number $N_d$ of depleted particles can give reliable information about dynamical properties and time evolution. Second, formula \eqref{depletionApproximate} is no longer valid close to $\lambda=1$. Therefore, it is conceivable that undercritical quantum breaking can take place if $\lambda$ is sufficiently close to $1$. Third, findings for $Q\gg N$ have meaning for systems with finite parameters and finite $\hbar$, independently of their limiting behavior. While these questions remain to be answered, it is important to emphasize that our NPM is a consistent quantum system for all values of $Q$ and our numerical analysis of time evolution is viable for all parameter choices.

\subsection{Overcritical regime}\label{sec:ModelOver}
Next, we shall consider the overcritical regime $\lambda > 1$. As is evident from \eq \eqref{depletionOvercritical}, the number of depleted particles scales with $N$ already for $Q=1$. Therefore, the inclusion of additional species does not change this finding. In order to obtain an estimate for the quantum break-time, we take into account the effect of species in the rate \eqref{rateOvercritical}:
\begin{equation} \label{rateOvercriticalSpecies}
		\Gamma = Q \lambda^2 + Q \sqrt{\lambda - 1} \Delta N\;.
	\end{equation}
	As in \eq \eqref{rateUndercriticalSpecies}, the initial rate is enhanced by a factor of $Q$. In order to decide which of the two terms dominates, we shall study what maximal value of $\Delta N$ could be obtained. If  the second term in \eq \eqref{rateOvercriticalSpecies} dominates, then $\Delta N$ is determined by the point when the instability disappears due to backreaction. As derived in \eq \eqref{turningPointOvercritical}, this leads to 
	\begin{equation}\label{deltaNInstability}
		\Delta N_{\text{inst.}}=N(\lambda-1)/\lambda \;.
	\end{equation}
	Crucially, $\Delta N_{\text{inst.}}$ is independent of $Q$. Therefore, one can expect that for sufficiently large $Q$, the first term in \eqref{rateOvercriticalSpecies}, corresponding to perturbative scatterings, becomes important. In this parameter regime, we use \eq \eqref{depletionSimplified} to obtain the estimate\footnote
	{First, we assume $N_d\ll Q\ll \lambda N$ to derive from \eq \eqref{depletionSimplified}:
		\begin{equation}
			\left(1-\lambda\left(1-\frac{N_d}{N}\right)\right) N_d = \frac{Q \lambda^2}{16} \;.
		\end{equation}
		We can solve this quadratic equation and obtain \eq \eqref{depletionEstimateManySpecies} for sufficiently small $\lambda-1$.
	}
	\begin{equation} \label{depletionEstimateManySpecies}
		\Delta N_{\text{pert.}} \approx \frac{\sqrt{Q N \lambda}}{4} \;.
	\end{equation}

Accordingly, we can try to estimate the quantum break-time. If the instability dominates, the enhancement of the rate by $Q$ in \eq \eqref{rateOvercriticalSpecies} leads to an additional factor of $1/Q$ in $t_q$ as compared to \eqref{breaktimeOvercritical}. The situation is more involved for sufficiently large $Q$. One could try to employ the result \eqref{breaktimeUndercriticalSpecies} of the undercritical regime but this would amount to completely neglecting the instability and the second term in \eq \eqref{rateOvercriticalSpecies}. Clearly, this cannot be the case, which prevents us from making an analytic statement for large $Q$. In total, depending on whether \eqref{deltaNInstability} or \eqref{depletionEstimateManySpecies} dominates, we estimate the quantum break-time as 
	\begin{equation}\label{breaktimeOvercriticalSpecies}
		t_q = \begin{cases}
			\frac{\ln N}{Q \sqrt{\lambda - 1}} & Q\ll 16 N \frac{(\lambda-1)^2}{\lambda^3}\\
			\boldsymbol{???}  &  Q\gg 16 N \frac{(\lambda-1)^2}{\lambda^3}
		\end{cases} \;,
	\end{equation}
	where only the numerical study will allow us to derive a conclusion about the regime of large $Q$. Moreover, we note that the interpretation of the regime $Q \gg N$ may cause problems. In analogy to our discussion of undercritical quantum breaking, it is possible that this parameter space leads to a quantum break-time that does not become infinite in the classical limit $\hbar\rightarrow0$.
 Finally, we must emphasize that all the above formulas only represent heuristic estimates. Whether or not they are valid will be determined by our subsequent numerical investigation.

\subsection{Critical regime}

In the critical case $\lambda = 1$, we will again use an extrapolation of the undercritical formula \eqref{depletionSimplified} to estimate the number of depleted particles. This yields
	\begin{equation} \label{depletionCriticalSpecies}
		N_d = \frac{Q^{2/3} N^{1/3}}{2^{4/3}} \left(1+O\left(\frac{1}{N^{2/9}}\right)\right) \;,
	\end{equation}
	\ie including $Q$ overcritical species enhances depletion by a factor of $Q^{2/3}$ (\cf \eq \eqref{depletionCritical}).
	We expect that the dependence $t_q\sim \sqrt{N}$ of the quantum break-time shown in \eq \eqref{breaktimeCritical} persists but we are unable to provide an analytic estimate regarding the influence of $Q$ on this scaling.

\section{Numerical study}
\label{sec:numeric}

Now we turn to the numerical study of the prototype model \eqref{simplifiedSpecies}. To this end, we employ the software \textit{TimeEvolver} \cite{Michel:2022kir}, which is capable of computing real-time evolution with adjustable error bound for generic quantum system of Hilbert space dimension up to $\sim 10^7$. Generically, we set in the following the bound on $||v_{\text{num}}(t) - v(t)||$ to be $10^{-6}$, where $v_{\text{num}}(t)$ is the output of \textit{TimeEvolver} and $v(t)$ corresponds to the exact state at time $t$. For selected parameters, corresponding to very large Hilbert spaces, we use as tolerance $10^{-4}$, mainly due to an effect of finite machine precision. We plot data obeying the stricter error bound with black dots while results obtained with a $10^{-4}$ tolerance are depicted in gray. Unless stated otherwise, we compute the expectation value of observables in intervals of $t_{\text{step}}=0.01$. All numerical data presented subsequently are publicly available in the companion Zenodo record \cite{zenodo2023}. The code to generate the data is available in the complementary github repository \cite{github2023}.
	
The feasibility of numerical simulation is mainly limited by Hilbert space dimension. In order to reduce the size of the problem, we introduce a maximal occupation $C$ for the $\hat{a}_k$-modes, $k \neq 0$. This is also empirically motivated since in some situations these modes are not highly occupied. However, we set value of $C$ so high that it does not qualitatively influence the behavior of the system. We can ensure that this property is fulfilled by increasing $C$ until its effect becomes negligible.

In the following analysis we will denote the expectation value of an operator $\hat{n}_k$  with the same symbol but without the hat: $\braket{\hat{n}_k} = n_k$. As initial state, we employ \eqref{initialState}, corresponding to all particles in the $\hat{a}_0$-mode, $\bra{\text{in}}\hat{n}_0 \ket{\text{in}} = N$. As discussed above, the advantage of using this state is that evaluating the expectation value $\braket{\hat{n}_0}$ is sufficient for determining departure from the classical solution. Quantum breaking occurs as soon as $\braket{\hat{n}_0}$ deviates significantly from $N$.

\subsection{Comparison of prototype model and periodic system}
A motivation for constructing our NPM \eqref{simplifiedSpecies} consists in its similarity to the PPM of \cite{Kanamoto2002} in its truncated version \eqref{periodicThreeModes}. In particular, we have shown that the results of Bogoliubov approximation coincide, \cf \eqs \eqref{periodicThreeModesBogo} and \eqref{simplifiedBogo}. 
We extend this comparison here to the full quantum Hamiltonians \eqref{periodicThreeModes} and \eqref{simplifiedSpecies} for finite total occupation number $N$. Obviously, those two systems are not identical and so we do not expect indistinguishable result. We shall see, however, that dynamics are qualitatively very similar.

To verify this claim and quantify the difference for finite $N$, we perform numerical simulations for both the under- and overcritical regime corresponding to the collective coupling values $\lambda_u = 0.8$ and $\lambda_o=1.2$, respectively. Moreover, we set the total particle number in both systems to be $N=10$. 
Since the truncated PPM \eqref{periodicThreeModes} features two modes with non-zero momentum, we set $Q = 2$ in the NPM. In both models we set the capacity $C$ equal to the number of particles to exclude any influence originating from finite maximum occupation numbers. We also set the symmetry-breaking coupling in the NPM to zero, $C_m=0$ to keep the symmetry between the $\hat{a}_1$ and $\hat{a}_2$ mode in analogy to the momentum conservation constraint in the PPM that forces $n_{-1}(t)=n_1(t)$. We note, however, that the analogy between the NPM and the PPM is not complete. While the PPM features terms coupling the non-zero modes, those are completely absent in the NPM when we set $C_m=0$. 
Our parameter choices are summarized in \eq \eqref{ParaComparisonPeriodic} for the PPM and in \eq \eqref{ParaComparisonSimplified} for NPM:
\begin{equation} \label{ParaComparisonPeriodic}
	\text{PPM  (\eq \eqref{periodicThreeModes}):}\qquad 	\lambda = \{0.8, \, 1.2\} \;, \qquad N=10 \;, \qquad C=10 \;,
\end{equation}

\begin{align} \label{ParaComparisonSimplified}
	\text{NPM  (\eq \eqref{simplifiedSpecies}):}\qquad 	&\lambda =\{0.8,\, 1.2\} \;, \qquad N=10 \;, \qquad C=10\;,\\
	&  Q=2 \;,  \qquad C_m=0  \;. \nonumber
\end{align}

\begin{figure}[h]
	\centering 
	\begin{subfigure}{0.95\textwidth}
		\centering
		\includegraphics[width=0.49\linewidth]{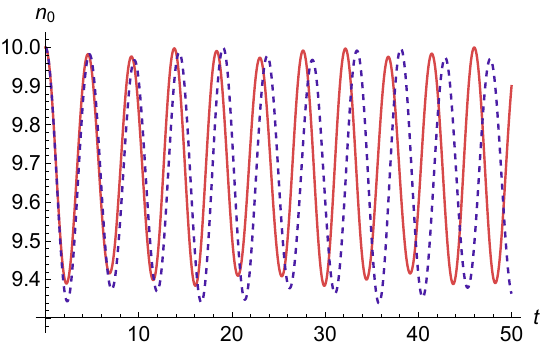}
		\hfill
		\includegraphics[width=0.49\linewidth]{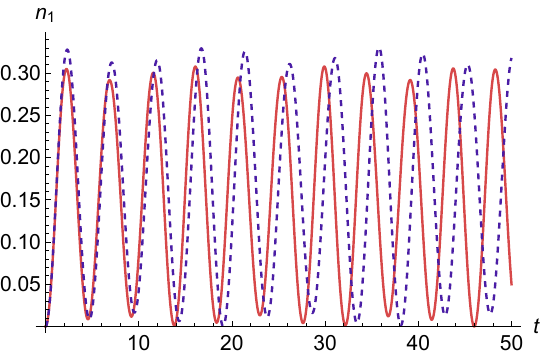}
		\caption{$\lambda=0.8$}
	\end{subfigure}
	\vspace{20pt}
	\begin{subfigure}{0.95\textwidth}
		\centering
		\includegraphics[width=0.49\linewidth]{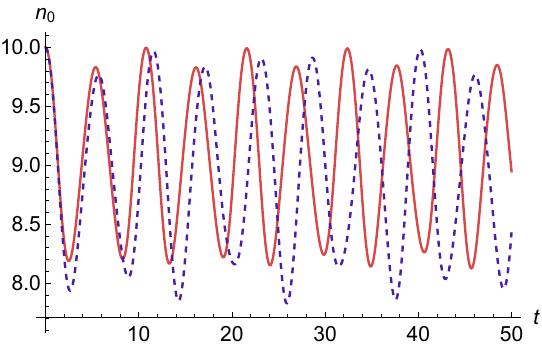}
		\includegraphics[width=0.49\linewidth]{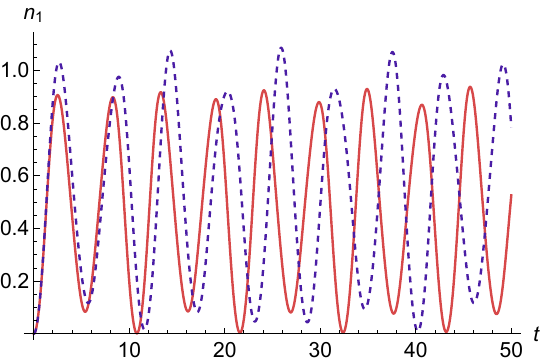}
		\caption{$\lambda=1.2$}
	\end{subfigure}
	\caption{Time evolution of expectation values of the particle number operator $n_0$ (left side) and $n_1$ (right side) with initial state \eqref{initialState} and parameter values set by \eqref{ParaComparisonPeriodic} and \eqref{ParaComparisonSimplified}. The upper row depicts the undercritical regime $\lambda_u=0.8$ while the lower row shows the evolution in the overcritical regime $\lambda_o=1.2$.
		The PPM \eqref{periodicThreeModes} is depicted with a dashed blue line while the NPM \eqref{simplifiedSpecies} is drawn in red. Both systems show similar dynamics in terms of frequency and amplitude.}
	\label{fig:CompWithPeriodic}
\end{figure}

The realtime evolution of the occupation numbers of the  $\hat{n}_0$- and  $\hat{n}_1$- operators for both systems are plotted in \fig \ref{fig:CompWithPeriodic}. Note that due to a shift symmetry and corresponding momentum conservation $n_1 = n_{-1}$ in the PPM. In contrast, $n_2$ in the NPM is generically only restricted by particle number conservation, such that $n_2 = N - n_0 - n_1$.  We observe that in both cases the systems exhibits very similar qualitative and quantitative behavior. In particular, frequency and amplitude are of the same order of magnitude. 

\subsection{Undercritical regime}
\label{sec:numericUnder}
Having shown that the truncated PPM \eqref{periodicThreeModes} and the NPM \eqref{simplifiedSpecies} exhibit very similar dynamics, we shall only consider the latter in the following. As discussed in sections \ref{sec:Periodic} and \ref{sec:Model}, this model does not exhibit quantum breaking in the undercritical regime without additional species, \ie for $Q=1$. This is because the system both only shows very small deviations from the initial state as well a strong tendency to return to it on the natural frequency of the system. 
We argued, however, in section \ref{sec:Model} that the introduction of additional species or decay channels can result in a strong deviation from the initial state without recurrence. 

\begin{figure}
	\centering 
	\begin{subfigure}{0.45\textwidth} 
		\includegraphics[scale=0.8]{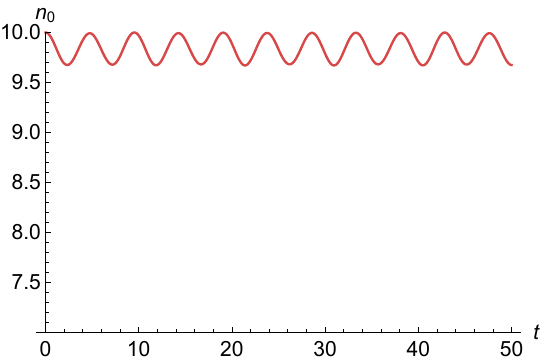}
		\caption{$Q=1$}
		\label{fig:QExamplesQ1}
	\end{subfigure}
	\hspace{0.05\textwidth}
	\vspace{40pt}
	\begin{subfigure}{0.45\textwidth} 
		\includegraphics[scale=0.8]{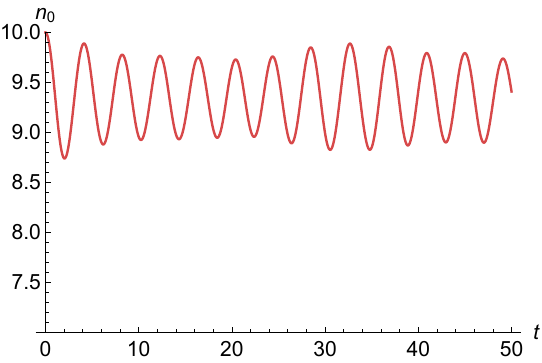}
		\caption{$Q=5$}
		\label{fig:QExamplesQ5}
	\end{subfigure}
	\begin{subfigure}{0.45\textwidth}
		\includegraphics[scale=0.8]{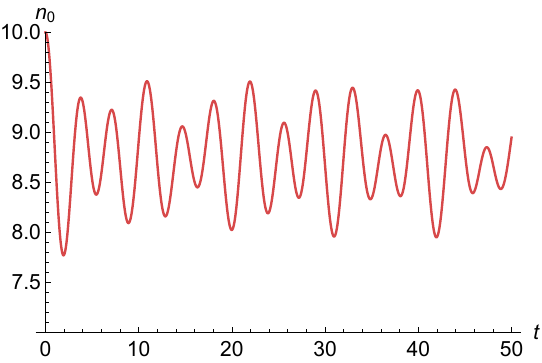}
		\caption{$Q=10$}
		\label{fig:QExamplesQ10}
	\end{subfigure}
	\hspace{0.05\textwidth}
	\begin{subfigure}{0.45\textwidth}
		\includegraphics[scale=0.8]{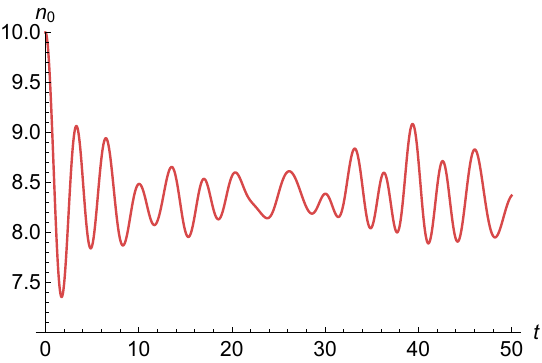}
		\caption{$Q=15$}
		\label{fig:QExamplesQ15}
	\end{subfigure}
	\caption{Time evolution of the initial state \eqref{initialState} for different values of $Q$. All other parameters are set by \eq \eqref{valuesUndercriticalExample}. The plots show the expectation value of $\hat{n}_0$ as function of time. While $Q=1$ leads to recurrence to the initial configuration, the phenomenon disappears for larger values of $Q$.}
	\label{fig:QExamples}
\end{figure}

In order to show that this is indeed the case, we perform numerical simulations for various values of $Q$. For breaking the symmetry between different modes, we also turn on the inter-species coupling $C_m$. The specific choices of parameters are given by
\begin{equation} \label{valuesUndercriticalExample}
	\lambda = 0.8 \;, \qquad N=10 \;, \qquad C_m = 0.1 \;, \qquad C=4  \;.
\end{equation}
We set $C<N$ this time but in the undercritical regime modes different than $n_0$ only get very sparsely occupied so that we do not expect the truncation of maximal occupation to have a significant impact. We will confirm this later in various other scans. For different values of $Q$, we display the expectation value of the $\hat{n}_0$ as a function of time in \fig \ref{fig:QExamples}. We observe that for $Q=1$, no quantum breaking takes places (see \fig \ref{fig:QExamplesQ1}), \ie $\braket{\hat{n}_0}$ always stays close to its initial value and returns to it on is free frequency timescale. 
However, this changes distinctly for sufficiently large $Q$. We observe that the expectation value of $\hat{n}_0$ deviates significantly from its initial value as soon as $Q \gtrsim 10$ (see \fig \ref{fig:QExamplesQ10}). Moreover, it never returns to the initial state. Thus, the explicit numerical study confirms that the NPM \eqref{simplifiedSpecies} exhibits quantum breaking without the presence of a classical instability.

After we confirmed that introducing an sufficient amount of additional modes can lead to quantum breaking, we shall next investigate quantitatively how the timescale of quantum breaking depends on the different parameters of the system. Most important will be the dependence on the total particle number $N$. 
Since the consideration in section \ref{sec:Model} are only valid for large $N$, we shall choose bigger values of $N$ as compared to \eq \eqref{valuesUndercriticalExample}. Correspondingly, $C_m$ needs to be sufficiently small in order not to change the qualitative behavior of the system (see appendix \ref{app:Cm}).
These considerations can be taken into account with the following choice of parameters
\begin{equation} \label{valuesUndercritical}
	\lambda = 0.8 \;, \qquad N=50 \;, \qquad C_m = 0.016 \;, \qquad C=4 \;, \qquad Q=10 \;.
\end{equation}

For the subsequent analysis, we define quantum breaking as follows: Over the range of parameters $X\in \{Q,N,\alpha,C_m\}$ that we consider, we simulate the dynamics of the system \eqref{simplifiedSpecies} with the initial state given by \eq \eqref{initialState} until $t_\text{max} =50$. We determine in each case the minimal relative occupation of the $\hat{n}_0$ mode,
\begin{equation}
	r_\text{min}(X) = \min_t \frac{\braket{\hat{n}_0(t)}}{N},
\end{equation}
where the dynamics and therefore the minimal occupation value depends on the parameters $X$ of the system considered.
To define a common threshold deviation from the initial occupation number of $\hat{a}_0$, we require it to be crossed for all values of $X$ in question. Correspondingly, we define quantum breaking as the timescale $t_q$ after which the relative occupation of the zero mode, $n_0/N$, falls for the first time below
\begin{equation} \label{threshold}
	b_{th} = 1 - (1-\max_X n_\text{min}(X)) \cdot 0.8 \;.
\end{equation}
Since not only the timescale but also the maximal amplitude will generally depend on the precise value of all parameters, this definition can only be meaningful over a finite interval. If this interval is chosen too wide, comparing the timescale for crossing a fixed threshold can become meaningless: A value of $	b_{th}$ that is close to the maximal amplitude in one case may correspond to only a small deviation from the initial state for a different choice of parameters. Therefore, the threshold should be chosen such that it is not much smaller than the minimal value of $n_0/N$ for all values of the parameter in the interval considered. In order to ensure that this condition is fulfilled, we present next to each scan also the a plot of the maximum relative depletion until $t_{\text{max}}$ and the corresponding threshold.

\begin{figure}
	\centering 
	\begin{subfigure}{0.45\textwidth}
		\includegraphics[scale=0.75]{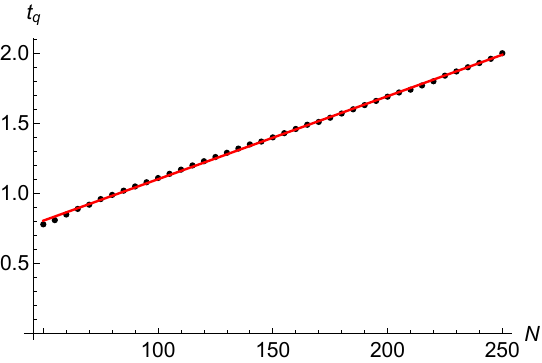}
		\caption{Quantum break-time as a function of $N$ depicted with black dotes. Linear fit drawn in red.}
	\end{subfigure}
	\hspace{0.05\textwidth}
	\begin{subfigure}{0.45\textwidth}
		\includegraphics[scale=0.75]{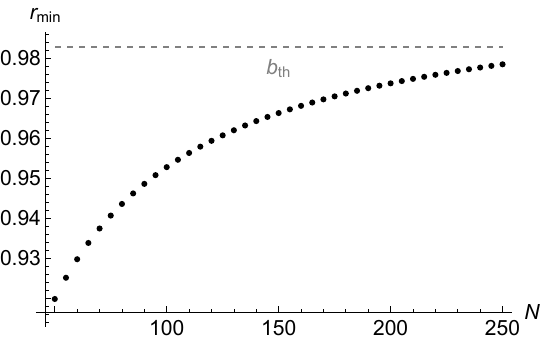}
		\caption{Minimum relative occupation. The threshold value for determining the break-time is indicated as dashed gray line.}
	\end{subfigure}
	\caption{$N$-Scan in the undercritical regime for $\lambda=0.8$ with remaining  parameters values set by \eq \eqref{valuesUndercritical}. The quantum break-time as a function of $N$ is shown on the left side and the corresponding threshold values for the amplitude are on the right side.
		We observe a linear dependence between the total particle number $N$ and the quantum break time $t_q$.}
	\label{fig:NScanUnder}
\end{figure}

\paragraph{Dependence on $N$} 
First, we investigate the scaling of the break-time with the total particle number $N$. 
To this end, we perform numerical simulations for the undercritical regime with parameter choices \eqref{valuesUndercritical} while varying the conserved particle number $N$. As shown in \eq \eqref{breaktimeUndercriticalSpecies}, analytic arguments suggest a linear scaling. This is confirmed in our simulation, as is evident from \fig \ref{fig:NScanUnder}, where we
scan over $ 50 \le N \le 250$ and following that perform a linear fit of the form
\begin{equation}\label{linearFit}
t_q =	m \cdot x + n \;,
\end{equation}
with fit parameters $m$, $n$ and $x=N$. Fitting the data yields the following values 
\begin{equation}\label{NlinFitParaUnder}
	m = 0.0059\;, \qquad  n = 0.51 \;.
\end{equation}
We also fit a polynomial function of the form 
\begin{equation}\label{polyFit}
t_q =	a \cdot x^c + b \;,
\end{equation}
with fit parameters $a, b, c$ in order to quantify the deviation from linearity.
Fitting the data  with a function of the form \eqref{polyFit} yields
\begin{equation}
	a = 0.0095\;, \qquad  b = 0.44\;, \qquad  c = 0.92 \;.
\end{equation}
The exponent $c$ close to $1$ indicates a linear relation between the particle number $N$ and the break-time $t_{q}$. 

As discussed above, we are required due to numerical limitation  to limit the maximal occupation of every mode. To check that we choose this capacity large enough to not have a qualitative effect on our findings, we repeated this analysis for higher $C=6$ in the appendix \ref{sec:appendixUnder} (see \fig \ref{fig:NScanUnder50_250_C6}). Moreover, we performed simulations for small $10 \le N \le 50$ in \fig \ref{fig:NScanUnder10_50} and large $1000 \le N \le 5000$ in \fig \ref{fig:NScanUnder1k_5k} as well as $C_m=0$ in \fig \ref{fig:NScanUnder50_250_Cm0}.
We observe that the linear behavior is maintained in all these examples. In particular, it does not depend qualitatively on a specific $N$ range or the inter-species coupling $C_m$. More details are shown in appendix \ref{sec:appendixUnder}.

\paragraph{Dependence on $Q$} 
Next, we investigate the scaling with the number of species $Q$. With the same parameter choice \eqref{valuesUndercritical}, we perform a scan for $Q \in \{15, 16, \dots , 29, 30\}$ and fit a polynomial function of the form \eq \eqref{polyFit}, where now $x=Q$. In agreement with \eq \eqref{breaktimeUndercriticalSpecies}, we observe (within numerical uncertainties) the following scaling
\begin{equation}
	t_q \sim Q^{-1.0} \;.
\end{equation}
The data as well as the fit details can be found in the appendix \ref{sec:appendixUnder}. Moreover, we perform a second $Q$-scan with lower particle number $N=10$ (and adjusted inter-species coupling $C_m$), which yields similar numerical values indicating the robustness of our results. We remark, however, that including smaller $Q$ in the fit increases the exponent. This can partly be attributed to threshold ambiguities but even besides that the fitting is not fully stable.
The figures for both fits can be found in the appendix, see \figs \ref{fig:QScanUnder} and \ref{fig:QScanUnder2}, respectively.

\paragraph{Dependence on $\lambda$}

Inspecting \eq \eqref{breaktimeUndercriticalSpecies} suggests an quadratic reciprocal scaling of the break-time with respect to the collective coupling $\lambda$. 
As explained at the beginning of this section, fitting over intervals in which the amplitude of $n_0$ changes significantly is not meaningful since no common threshold for defining quantum breaking can be found. 
Therefore, we split the fit in two intervals and fit over the regimes $0.2\le \lambda \le 0.5$ and $0.5\le \lambda \le0.9$ separately. We obtain following values
\begin{equation}
	t_q \sim \begin{cases}
		\lambda^{-1.8} &  0.2 \le \lambda \le 0.5 \;,\\
		\lambda^{-2.2} & 0.5 \le \lambda \le 0.9 \;,
	\end{cases}
\label{scalingLambdaUndercritical}
\end{equation}
which both lie near the analytically expected value of $c=-2$. Details and the corresponding \figs \ref{fig:LambdaScanUnderSmall} and \ref{fig:LambdaScanUnderLarge} can be found again in the appendix.We remark that we increase the sampling step to $t_{\text{step}}= 0.001$ for this scan as well as for the subsequent investigation of the dependence on $C_m$.

\paragraph{Dependence on $C_m$} 
As remaining parameter we investigate the dependence on the interspecies coupling $C_m$. This parameter is not present in the PPM discussed in section \ref{sec:Periodic}, where the interaction strength between different momentum modes is determined by the overall coupling constant $\alpha$. We usually set $C_m$ large enough to break the symmetry between the different species appreciably but at the same time small enough not to have a large influence on the dynamics. As is evident from \fig \ref{fig:CmScanUnder} in the appendix, our parameter choice fulfills this condition and the break-time exhibits only marginal sensitivity towards $C_m$. 

\subsection{Overcritical regime with few species}
\label{sec:numericOverFew}
After analyzing the undercritical regime we next investigate the overcritical phase with $\lambda >1$. 
As discussed in section \ref{sec:ModelOver}, we expect that the overcritical phase can be divided in two distinct regimes depending on the number of modes $Q$ in comparison with the total particle number $N$ and collective coupling $\lambda$. 
In this section we will first explore the case when there are only few species available to offload occupation numbers with respect to the total number of transferred particles, corresponding to the first line in \eq \eqref{breaktimeOvercriticalSpecies}. 

\begin{figure}
	\centering 
	\begin{subfigure}{0.45\textwidth}
		\includegraphics[scale=0.8]{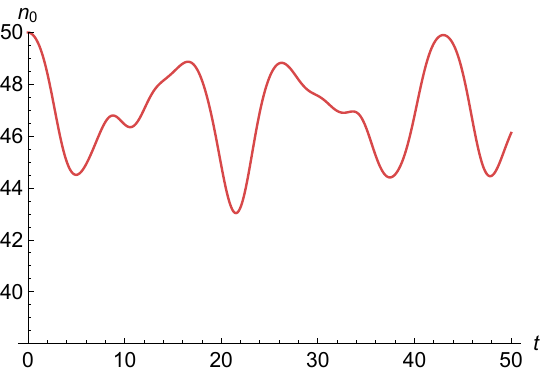}
		\caption{$Q=1$}
	\end{subfigure}
	\hspace{0.05\textwidth}
	\vspace{40pt}
	\begin{subfigure}{0.45\textwidth}
		\includegraphics[scale=0.8]{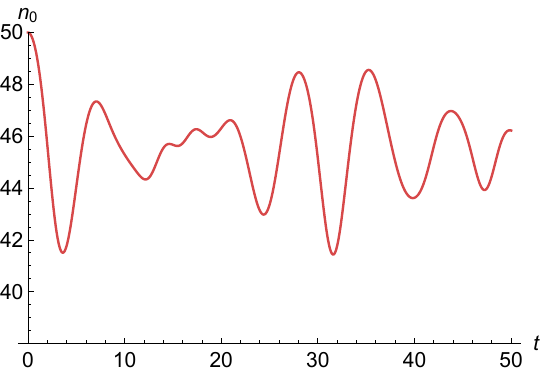}
		\caption{$Q=3$}
	\end{subfigure}
	\begin{subfigure}{0.45\textwidth}
		\includegraphics[scale=0.8]{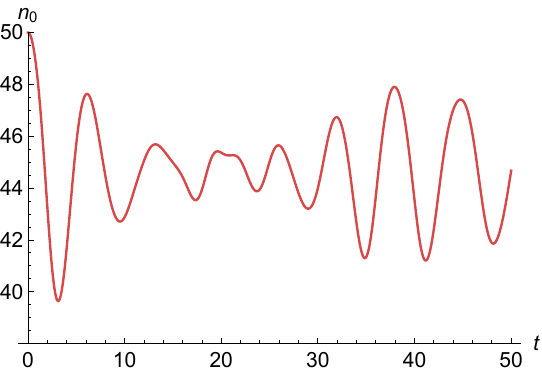}
		\caption{$Q=5$}
	\end{subfigure}
	\hspace{0.05\textwidth}
	\begin{subfigure}{0.45\textwidth}
		\includegraphics[scale=0.8]{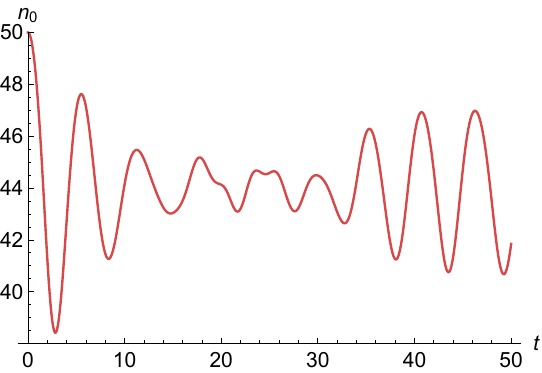}
		\caption{$Q=7$}
		\label{fig:QExamplesOverQ7}
	\end{subfigure}
	\caption{Time evolution of the initial state \eqref{initialState} for different values of $Q$ in the overcritical regime $\lambda =1.2$ for small $Q$. All other parameters are set by \eq \eqref{valuesOvercriticalFew}. The plots show the expectation value of $\hat{n}_0$ as function of time. While $Q=1$ already leads to quantum breaking, increasing $Q$ enhances this effect.}
	\label{fig:QExamplesOver}
\end{figure}

It is known that due to the development of an instability the overcritical regime exhibits quantum breaking even without the introduction of additional species \cite{Dvali:2013vxa}. Nevertheless, we first want to give a qualitative picture of how the dynamic is influenced by the introduction of more decay channels. To this end, we again show the realtime evolution of the expectation value of the number operator $\hat{n}_0$ for different values of $Q$. The parameters are set by
\begin{equation} \label{valuesOvercriticalFew}
	\lambda = 1.2 \;, \qquad N=50 \;, \qquad C_m = 0.016 \;, \qquad C=50 \;, \qquad Q=3 \;.
\end{equation}
This choice fulfills the heuristic criterion shown in \eq \eqref{breaktimeOvercriticalSpecies}, \ie $Q\ll 16 N (\lambda-1)^2/\lambda^3$. Since depletion is much stronger in the overcritical regime, we do not limit the occupation of the modes with $k\geq 1$ in order not to spoil our findings by finite capacity effects. The results of time evolution for different values of $Q$ are displayed in \fig \ref{fig:QExamplesOver}.
We observe that the system exhibits quantum breaking already for $Q=1$ in contrast to the undercritical regime, where breaking could only be realized by the introduction of a sufficient number of species. This is in line with previous results on the PPM, in which breaking was only observed in the overcritical regime \cite{Dvali:2013vxa,Zell:2019vpe}. However, although quantum breaking is already possible in the absence of additional species for $\lambda>1$, introducing more channels still increases the amplitude and shortens the timescale. 

Next, we investigate the system quantitatively and scan again over the different parameters, where in each case the remaining parameters are set by \eq \eqref{valuesOvercriticalFew}.
In addition to a polynomial fit function \eqref{polyFit}, we will also use a logarithm function of the form
\begin{equation}\label{logfit}
t_q =	p \cdot \log(x) + q \;,
\end{equation}
with fit parameters $q$ and $p$. 

\begin{figure}
	\centering 
	\begin{subfigure}{0.45\textwidth}
		\includegraphics[scale=0.75]{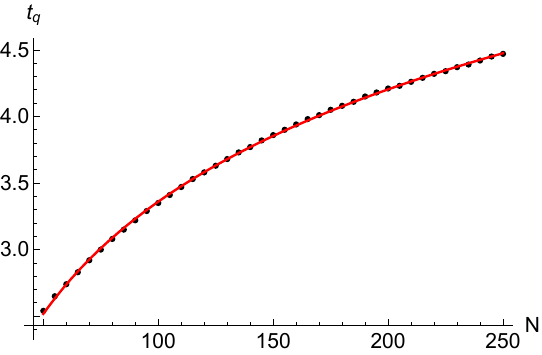}
		\caption{Quantum break-time as a function of $N$ depicted with black dotes. Logarithmic fit drawn in red.\vspace{20pt}}
	\end{subfigure}
	\hspace{0.05\textwidth}
	\begin{subfigure}{0.45\textwidth}
		\includegraphics[scale=0.75]{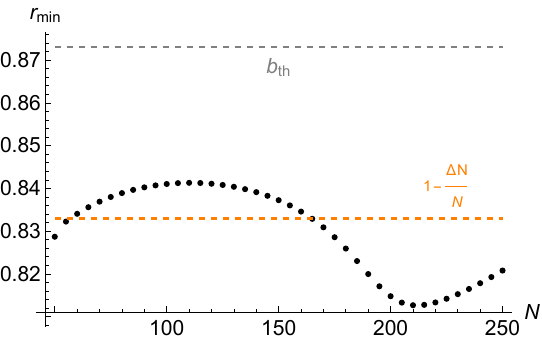}
		\caption{Minimum relative occupation. The threshold value for determining the break-time is indicated as dashed gray line. Turning point due to backreaction to the collective coupling in orange (see \eq \eqref{turningPointOvercritical}).}
	\end{subfigure}
	\caption{$N$-Scan in the overcritical regime for $\lambda=1.2$ with remaining  parameters values set by \eq \eqref{valuesOvercriticalFew}. Quantum break-time as a function of $N$ on the left side and the corresponding threshold values for the amplitude on the right side. 
		We observe a logarithmic dependence between the total particle number $N$ and the quantum break time $t_q$.}
	\label{fig:NScanOverFew}
\end{figure}

\paragraph{Dependence on $N$} 
First we study again the dependence of the break-time on the total particle number $N$. We show the data and fit for the value range $ 50 \le N \le 250$ in  \fig \ref{fig:NScanOverFew}.	
In agreement with \eq \eqref{breaktimeOvercriticalSpecies}, we observe a logarithmic scaling with the fit parameters
\begin{equation}
	p =  1.22\;, \qquad  q =-2.24 \;.
\end{equation}
Again we perform additional scans to check the generality of our findings. First we increase the numerical parameter $C$ to verify that we set the capacity high enough to not spoil our findings. Moreover, we also check that the interspecies coupling has no significant quantitative influence by repeating the analysis with $C_m=0$. Both checks yields almost identical results. The difference to the fit depicted in \fig \ref{fig:NScanOverFew} is minuscule. The data, fits and plots can again be found in appendix \ref{sec:appendixOverFew}, see \fig \ref{fig:NScanOverFewC60} for $C=60$ and see \fig \ref{fig:NScanOverFewCm0} for $C_m=0$.

\paragraph{Dependence on $Q$}
After finding the previously known scaling with $\ln N$, we now turn to the other parameters. We start again with the dependence on the number of species $Q$. We use \eq \eqref{valuesOvercriticalFew} as base values and perform a scan over increasing $Q$. Unfortunately, due to the exponential scaling of the Hilbert space with $Q$ and the corresponding increase in demand of computational resources, we are very limited in the size of $Q$ we can simulate. 
Correspondingly, we perform a scan over $2 \le Q \le 8$.\footnote
{Note that we excluded $Q=1$ since in this case a larger amplitude of $n_0$ was reached after the system recurred closely to the initial state.}
 The result is plotted in \fig \ref{fig:QScanOverFew} and can be found with the corresponding details in the appendix. Selected realtime plots have already been shown in the beginning of this section in \fig \ref{fig:QExamplesOver}. In summary we find the following dependence
\begin{equation}
	t_q \sim Q^{-0.84} \;,
\end{equation}
which is close the expected value $c=-1$ of \eq \eqref{breaktimeOvercriticalSpecies}. We want to stress, however, that the fit is not fully stable and the exponent of $Q$ can  vary depending on the fit range.

\paragraph{Dependence on $\lambda$}
Next, we scan over the collective coupling $\lambda$, where we limit ourselves to the interval $1.2\le \lambda \le1.6$. In this regime, we expect the scaling $\frac{1}{\sqrt{\lambda -1}}$ as shown in \eq \eqref{breaktimeOvercriticalSpecies}. 
We therefore perform a fit of the form
\begin{equation}\label{fitShiftedPoly}
	a \cdot (x - d)^c \;,
\end{equation}
which yields
\begin{equation} \label{lambdaFitOvercriticalFew}
	t_q \sim \frac{1}{(\lambda-0.88)^{0.49}} \;,
\end{equation}
 in very good agreement with the analytic expectations. For details and figures we refer the reader to \fig \ref{fig:lambdaScanOverFew2} in the appendix.

\paragraph{Dependence on $C_m$}
In analogy to the undercritical case, we want to verify that $C_m$ does not play a significant role in the dynamics, as long as the bounds of appendix \ref{app:Cm} are maintained. For this purpose, we scan over $ 0.01 \le C_m \le 0.09$. In this interval, the change of the break-time is only of order $\sim 10^{-3}$, as is evident from \fig \ref{fig:CmScanOverFew} in the appendix.

\subsection{Overcritical regime with many species}
\label{sec:numericOverMany}
Finally, we want to study the overcritical regime with a large number $Q$ of modes. We implement this by the following numerical choice of parameters
\begin{equation} \label{valuesOvercriticalMany}
	\lambda = 1.2 \;, \qquad N=10 \;, \qquad C_m = 0.08 \;, \qquad C=16 \;, \qquad Q=10 \;,
\end{equation}
which fulfill the heuristic criterion shown in \eq \eqref{breaktimeOvercriticalSpecies}, \ie $Q\gg 16 N (\lambda-1)^2/\lambda^3$.
Moreover, we note that we set $C>N$ in order to accommodate for higher $N$ in the $N$-scan. Due to computational limitations, large $Q$ requires us to use this relatively small $C$ which can have a quantitative influence on the time evolution of the system. However, we will again verify that the results are still qualitatively independent of the specific choice of $C$. Exploring the second overcritical regime is especially computationally intensive since the Hilbert space dimension scales exponentially with $Q$ and $Q$ needs to be sufficiently large as compared to $N$ over the whole scan range. On the other hand, $N$ cannot be to small to avoid finite size effects scaling as $1/N$.

\begin{figure}
	\centering 
	\begin{subfigure}{0.45\textwidth}
		\includegraphics[scale=0.8]{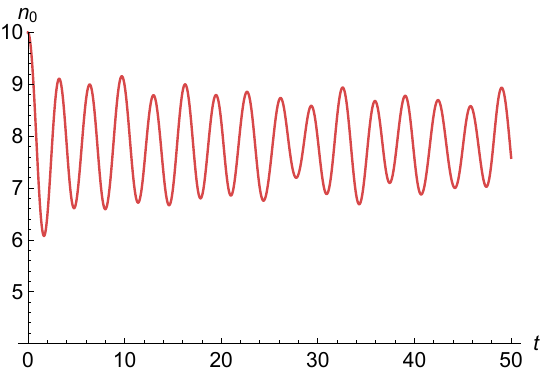}
		\caption{$Q=10$}
	\end{subfigure}
	\hspace{0.05\textwidth}
	\vspace{40pt}
	\begin{subfigure}{0.45\textwidth}
		\includegraphics[scale=0.8]{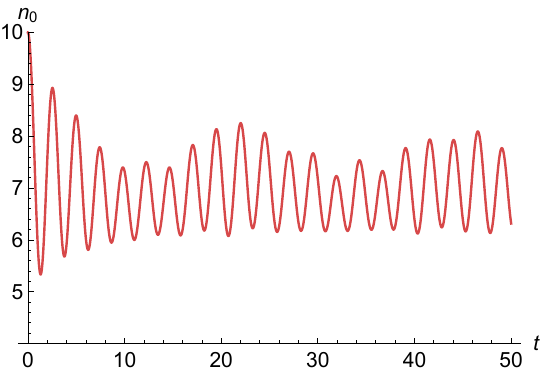}
		\caption{$Q=20$}
	\end{subfigure}
	\begin{subfigure}{0.45\textwidth}
		\includegraphics[scale=0.8]{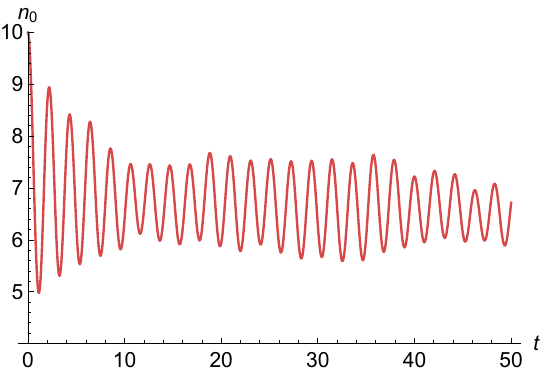}
		\caption{$Q=30$}
	\end{subfigure}
	\hspace{0.05\textwidth}
	\begin{subfigure}{0.45\textwidth}
		\includegraphics[scale=0.8]{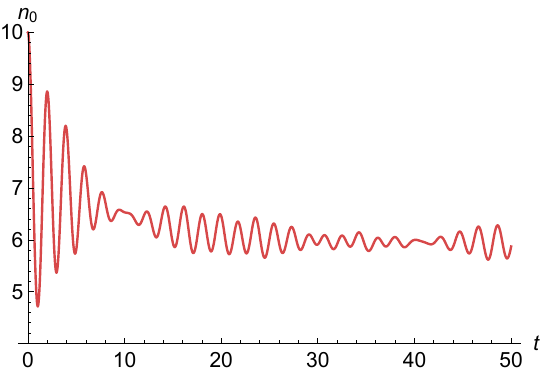}
		\caption{$Q=40$}
		\label{fig:QExamplesOverMany40}
	\end{subfigure}
	\caption{Time evolution of the initial state \eqref{initialState} for different values of $Q$ in the overcritical regime $\lambda =1.2$ for large $Q$. All other parameters are set by \eq \eqref{valuesOvercriticalMany}. The plots show the expectation value of $\hat{n}_0$ as function of time. Increasing $Q$ enhances the breaking effect.}
	\label{fig:QExamplesOverMany}
\end{figure}

We start again by showing selected real time plots for various numbers of species $Q$ in \fig \ref{fig:QExamplesOverMany}. Since we start in the overcritical regime (and have large $Q$) all plots exhibit quantum breaking. Similar to the case of few $Q$ discussed in the last section, increasing the number of channels enhances the breaking effect significantly. This effect is particularly strong in \fig \ref{fig:QExamplesOverMany40}.

\begin{figure}
	\centering 
	\begin{subfigure}{0.45\textwidth}
		\includegraphics[scale=0.75]{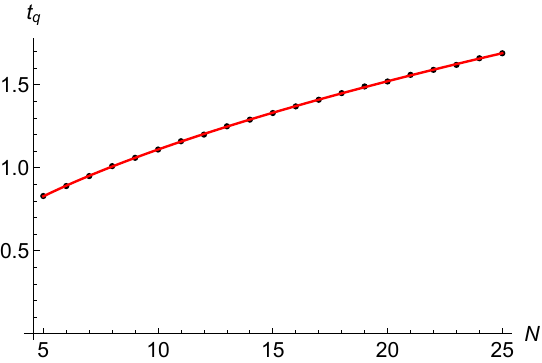}
		\caption{Quantum break-time as a function of $N$ depicted with black dotes. Polynomial fit drawn in red.}
	\end{subfigure}
	\hspace{0.05\textwidth}
	\begin{subfigure}{0.45\textwidth}
		\includegraphics[scale=0.75]{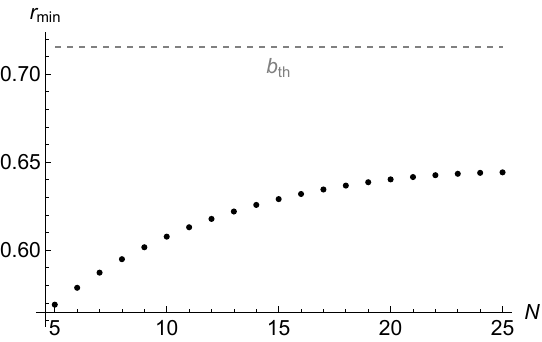}
		\caption{Maximum relative depletion. The threshold value for determining the break-time is indicated as dashed red line.}
	\end{subfigure}
	\caption{$N$-Scan in the overcritical regime for $\lambda=1.2$ and many species with remaining  parameters values set by \eq \eqref{valuesOvercriticalMany}. Quantum break-time as a function of $N$ on the left side and the corresponding threshold values for the amplitude on the right side. 
	We observe that the quantum break-time $t_q$ scales with the square root of the total particle number $N$.}
	\label{fig:NScanOverMany}
\end{figure}

\paragraph{Dependence on $N$}
We start again with the dependence on $N$. As discussed in section \ref{sec:ModelOver}, we do not have a clear analytical expectations for this regime. 
We consider the parameter range $5 \le N \le 25$ and show the corresponding data in \fig \ref{fig:NScanOverMany}. Performing again a polynomial fit, we find following values
\begin{equation}
	a =0.26\;, \qquad  b = 0.21\;, \qquad  c = 0.54 \;.
\end{equation}
Therefore, we get
\begin{equation}\label{NscalingOverMany}
	t_q \sim N^{0.54} \;.
\end{equation}
The numerical results suggest a dependence on the square root of the particle number (\cf \eq \eqref{rootScalingN}). This is a novel regime distinct from the linear scaling in the undercritical phase and the logarithmic dependency in the overcritical phase with few species. However, we have to stress that the true value for the exponent might differ significantly from \eq \eqref{NscalingOverMany}. On the one hand, we both expect finite size effect due to a relatively low particle number $N$. On the other hand, the heuristic conditions for being in the regime of large species (see \eq \eqref{breaktimeOvercriticalSpecies}) is only marginally fulfilled close to $N \approx 25$ when $Q=10$.
Again we perform two additional consistency checks by performing $N$-scans for $C_m=0$ and higher capacity $C=20$. These can be found in the appendix \ref{sec:appendixOverMany}. The numerical values for the exponent in those scans are close to \eq \eqref{NscalingOverMany}, yielding further indications for the scaling $t_q \sim \sqrt{N}$.

\paragraph{Dependence on $Q$} 
Next, we perform a scan over $15 \le Q \le 40$. More details can again be found in the appendix and the result is displayed in \fig \ref{fig:QScanOverMany}. Fitting the polynomial function \eqref{polyFit}, we observe the dependence
\begin{equation}
	t_q \sim Q^{-0.67} \;.
\end{equation}
This finding could indicate a scaling with $Q^{-2/3}$, which, however, remains to be verified. 

\paragraph{Dependence on $\lambda$} 
To determine the $\lambda$ dependence, we compute the time evolution for several values in the interval $\lambda \in [1.1,1.3]$. Again, we do not have a clear expectation and so we perform a shifted polynomial fit of the form \eq \eqref{fitShiftedPoly}. This yields 
\begin{eqnarray}\label{fitLambdaOverMany1}
	t_q \sim \frac{1}{(\lambda-0.66)^{0.6}} \;,
\end{eqnarray}
However, it is also possible to fit the data with a polynomial function of the form \eq \eqref{polyFit}. In this case the break-time scales as 
\begin{eqnarray}\label{fitLambdaOverMany2}
	t_q \sim \lambda^{-2.59} \;.
\end{eqnarray}
We observe behavior similar to the undercritical case (see \eq \eqref{scalingLambdaUndercritical}) continuing the tendency of ever more negative exponent for larger $\lambda$. At the same time the data also indicate a divergent property at a critical point similar to the overcritical case with only few available species (see \eq \eqref{lambdaFitOvercriticalFew}). 
At this point, we cannot conclusively determine the true scaling behavior of the collective coupling. Both \eq \eqref{fitLambdaOverMany1} and \eq \eqref{fitLambdaOverMany2} describe the data equally well. 
Details concerning the fits and the resulting figures can be found in the appendix (see \fig \ref{fig:ScansOverManyLambda}). 

\paragraph{Dependence on $C_m$} 
The final scan we perform is again over the interspecies coupling $C_m$. This time we vary its value such that $0 \le C_m \le 0.5$. Although the interval is relatively large, we observe again only marginal influence on the break-time $t_q$. The resulting \fig \ref{fig:CmScanOverManyW} can be found in the  appendix.

\subsection{Critical interpolation}
\label{sec:numericInterpolation}
So far, we have investigated the dependence of the quantum break-time on the total particle $N$ for different regimes. We found a linear dependence in the undercritical regime $\lambda <1$ and a logarithmic scaling for the overcritical case $\lambda >1$ corresponding to a classical instability. In this final section, we study the transition between these two regimes. We will do this by performing several $N$-scans for different values of the collective coupling in the interval $0.8 \le \lambda \le 1.2$.
The threshold amplitude is determined for each $\lambda$ individually. We then perform fits to determine the interpolation between the previously observed linear and logarithmic scalings.

\begin{figure}
	\centering 
	\begin{subfigure}{1.0\textwidth}
		\centering
		\includegraphics[scale=1.0]{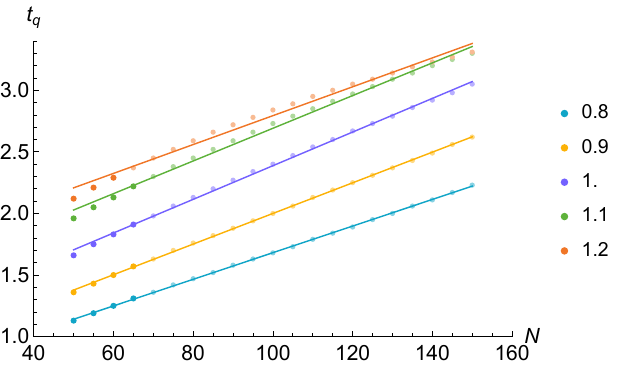}
		\caption{Linear fits.}
		\label{fig:InterpolationLin}
	\end{subfigure}
	\hspace{0.05\textwidth}
	\begin{subfigure}{1.0\textwidth}
		\centering
		\includegraphics[scale=1.0]{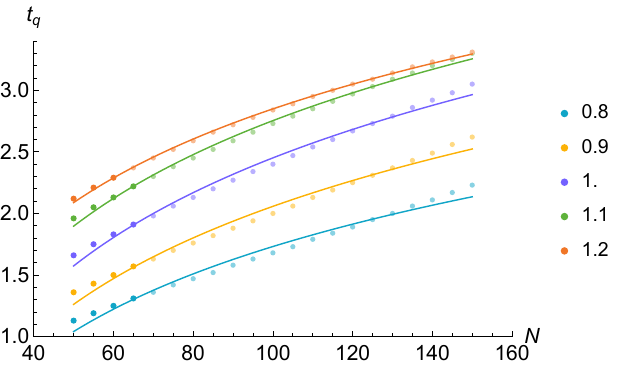}
		\caption{Logarithmic fits.}
		\label{fig:InterpolationLog}
	\end{subfigure}
	\caption{Quantum break-time as a function of $N$ for various $\lambda$ interpolating between the undercritical regime, $\lambda <1$, and the overcritical phase, $\lambda >1$. Linear fits in \fig \ref{fig:InterpolationLin} and logarithmic fits in \fig \ref{fig:InterpolationLog} on the same data. While for low collective coupling $\lambda$ linear fits perform better, a $\log N$ dependence describes the scaling more accurately for higher $\lambda$. 
		The break-time interpolates continuously between a linear regime and a logarithmic one. Note that the increase in the break-time for higher $\lambda$ is caused by an adjusted threshold for each $\lambda$.}
	\label{fig:Interpolation}
\end{figure}

First, we have to find a set of values for the parameters that are physically meaningful throughout the phase transition and obey numerical limitations. Specifically, the implementation of quantum breaking in the undercritical regime requires the introduction of a sufficient number of additional species $Q$. Moreover, we also require $N$ to be sufficiently large in order to avoid finite size effects. However, the overcritical regime also needs a large capacity $C\sim N$ in order not to spoil the logarithmic scaling. Ideally one would like to choose all these parameters sufficiently high, however, the exponential scaling of Hilbert space dimension makes this unfeasible. Therefore, we have to take a compromise here and we will vary the particle number within $50 \le N \le 150$ with the other parameters set by
\begin{equation} \label{valuesInterpolation}
	\qquad C_m = 0.01 \;, \qquad C=44 \;, \qquad Q=5 \;.
\end{equation}
Note that $C=44$ is already so small that it starts to influence dynamics close to $N=150$  in the overcritical regime. Therefore, quantitative results should be interpreted with care.
However, the qualitative transition from the undercritical to the overcritical regime can still be captured. 
In order to visualize the transition, we plot the break-time and fit a linear function in \fig \ref{fig:InterpolationLin} as well as a logarithm in \fig \ref{fig:InterpolationLog}. We conclude that while a linear fit describes the data very precisely for small $\lambda \lesssim 0.9$, it becomes increasingly less accurate for higher $\lambda$ and does not provide a meaningful description for $\lambda \gtrsim 1.1$. Conversely, a logarithmic fit does not capture the break-time dependence on $N$ in the undercritical phase but becomes increasingly precise in the overcritical regime. 

\begin{figure}
	\centering
	\includegraphics[scale=1.0]{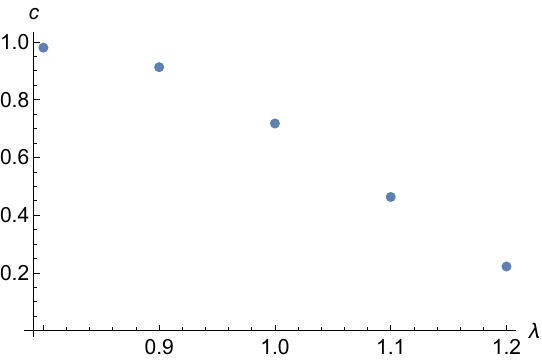}
	\caption{Numerical values of exponent fit parameter $c$ of a polynomial fit (see \eq \eqref{polyFit}).}
	\label{fig:fitParameterc}
\end{figure}

In addition to the qualitative behavior shown in \fig \ref{fig:Interpolation}, we provide also a quantitative analysis of the transition between both phases. On top of a linear and $\log$ fit shown in  \figs \ref{fig:InterpolationLin} and  \ref{fig:InterpolationLog}, respectively, we also perform a polynomial fit of the form \eq \eqref{polyFit} for each $\lambda$ and plot the resulting exponent $c$ in \fig \ref{fig:fitParameterc}. For $\lambda=0.8$, the numerical value of the exponent is close to $c=1$, in accordance with a linear scaling. In contrast, $c$ is much smaller for $\lambda=1.2$, which indicates a logarithmic dependence.\footnote
{For sufficiently small $c$, one can approximate
	\begin{equation}
		a x^c + b = a\left(1 + c \ln x + O\left((c \ln x)^2\right)\right) + b \;.
	\end{equation}
	Thus, only keeping the term linear in $\ln x$ is a good approximation as long as $c \ln x \ll1$ for the largest value of $x$ under consideration. Since we scan up to $N=150$, the critical value of $c$ is $1/\ln 150 \sim 0.2$.}
When increasing $\lambda$ from $0.8$ to $1.2$, the observed value of $c$ decreases monotonously. This leads to a smooth interpolation between the linear and logarithmic scalings.

We want to note again that the threshold in \fig \ref{fig:Interpolation} was set for each $\lambda$ individually. Since the threshold $b_{th}$ as defined in \eq \eqref{threshold} deviates more significantly from $1$ for higher $\lambda$, the break-time appears to increase. There is specifically no inconsistency with the $\lambda$-scaling found in the previous chapter that were based on a fixed threshold for the entire $\lambda$ range. We comment on this point in more detail in appendix \ref{sec:appendixCritInterpolation}, where we also perform a fit in which the same threshold is used for all values of $\lambda$. Finally, we want to emphasize that part of the deviation from a perfect logarithmic relation for $\lambda=1.2 $ originates from setting the maximal occupation per mode $C<N$.

\clearpage
\section{Conclusion}
\label{sec:Conclusion}

\subsection{Summary of findings}
A classical description generically represents an approximation of a more fundamental theory that includes quantum effects. In this paper, we have investigated the quantum break-time $t_q$, after which the classical approximation breaks down. To this end, we have developed further the system \cite{Kanamoto2002,Dvali:2013vxa} to construct a new prototype model (NPM) defined in \eq \eqref{simplifiedSpecies}. It can be characterized by two parameters in addition to the particle number $N$. The first one is the collective coupling $\lambda = \alpha N$, where $\alpha$ sets the strength of elementary quantum interactions. We can normalize $\lambda$ such that the system exhibits a classical instability for $\lambda > 1$. The other parameter corresponds to the number $Q$ of dynamically accessible quantum modes. 

After deriving analytic estimates, we have employed the software \textit{TimeEvolver} \cite{Michel:2022kir} to numerically compute full quantum dynamics. While detailed results are displayed in the main part, a simplified summary of our findings reads as follows:
In the generic case in which the system is classically stable, quantum breaking only takes place if $Q$ is sufficiently large. In this case, the dependence on the particle number is linear, $t_q \sim N$. In contrast, the regime $\lambda>1$, in which a classical instability is present, causes a breakdown of the classical description for all values of $Q$. If the number of modes is not too large, $Q\ll N$, we get the timescale $t_q\sim \ln N$ whereas a large $Q$ implies
\begin{equation} \label{scalingSquareRootConclusions}
	t_q \sim \sqrt{N} \;.
\end{equation}
Finally we have investigated the transition region $\lambda \approx 1$ and found 
\begin{equation}  \label{scalingPolynomialConclusions}
t_q\sim N^\gamma \;,
\end{equation}
where $0<\gamma<1$. The scaling \eqref{scalingPolynomialConclusions} smoothly interpolates between the linear scaling corresponding to $\lambda \ll 1$ and the logarithmic dependence caused by $\lambda\gg1$. In particular, we also observed the scaling \eqref{scalingSquareRootConclusions} close to the critical value $\lambda=1$.

Moreover displaying the dependencies on $Q$ and $\lambda$, we can schematically summarize the above discussion,
\begin{equation} \label{summaryScalings}
	t_q \sim \begin{cases}
		\displaystyle \frac{N}{Q \lambda^2}  & \lambda \ll 1\\
	\displaystyle	\frac{\ln N}{Q \sqrt{\lambda -1}} & \lambda \gg 1\ \&\ Q\ll N \\
		\displaystyle	\sqrt{N} f_1(Q,\lambda) & \lambda \gg 1\ \&\ Q\gg N \\
		\displaystyle  N^\gamma f_2(Q,\lambda) & \text{otherwise}
	\end{cases} \;,
\end{equation}
where $0<\gamma<1$ and $f_1(Q,\lambda)$, $f_2(Q,\lambda)$ are functions of $Q$ and $\lambda$. 
For fixed $Q$, all scalings shown in \eq \eqref{summaryScalings} are compatible with the classical limit.  In terms of classical quantities, namely the total energy $E$ and the elementary frequency $\omega$ of (free) oscillations, the number of particles scales as $N\sim E/(\hbar \omega)$ (see \cite{Dvali:2017eba,Dvali:2017ruz}). Consequently, the classical limit $\hbar$ leads to $N\rightarrow \infty$ and $t_q\rightarrow \infty$, \ie the classical approximation never breaks down.

However, there is an important caveat regarding quantum breaking in the undercritical regime. We have found indications that in this case quantum breaking is only possible if $Q\gtrsim N$. As is evident from the first line of \eq \eqref{summaryScalings}, this would imply that $1>t_q\sim \hbar^0$, corresponding to an essentially instantaneous breakdown of the classical approximation. Thus, no classical description exists to begin with and so it becomes meaningless to search for its breakdown. Should this finding be confirmed, then the following far-reaching statement about our model would be true: \textit{If there is no classical instability and a viable classical description exists at some initial time, then this classical description remains valid indefinitely.} Evidently, such a finding could shed light on the long-standing question why a complete deviation from the classical approximation is such a rarely observed phenomenon.

Several questions remain to be answered. While we have provided heuristic derivations for the first two lines of \eq \eqref{summaryScalings}, such an analytic argument is missing in the other cases. In particular, it would be interesting to better understand how the scaling $t_q\sim \sqrt{N}$ arises. Moreover, we have only established when the classical description ceases to be valid but not what exactly happens at this point. As a next step, we are currently investigating signatures of a deviation from the classical approximation, \eg related to higher point functions \cite{ToAppear}. Most of all, it remains to be determined if our results also apply to other systems. This is especially important in the regimes that we have for the first time observed dynamically, such as the scaling \eqref{scalingSquareRootConclusions}, as well as for the question if quantum breaking may become impossible in the absence of a classical instability. Evidently, systems that are experimentally realizable are of special interest\footnote
{For example, the Hamiltonian \eq \eqref{simplifiedSpecies} can also be thought of as originating from a multi-component or spinor BEC \cite{ohmi1998, ho1998}.	In a spinor condensate, the various $k$-modes could label different magnetic sub-levels $m$ where $-J \le m \le J$ with total angular momentum $J$. In the absence of an external magnetic field, the different spin modes have identical kinetic terms while their interactions can be modeled by the same term structure as in \eq \eqref{simplifiedSpecies} in the condensate phase as long as only the lowest momentum mode is excitable (see \cite{kawaguchi2012} for a review of spinor BECs). Spinor condensates with a similar number of (nearly)-degenerate modes as considered in this study are already experimentally accessible, \eg for $J=8$ \cite{lu2011} or for $J=6$ \cite{aikawa2012}.}.

\subsection{Outlook: connection to gravity}
In this work, we have developed further the system \cite{Dvali:2013vxa}, which originated from the quantum N-portrait of a black hole \cite{Dvali:2011aa}. In this microscopic model that we reviewed in section \ref{ssec:analogue}, a black hole of mass $M$ consists of $N=G_N M^2/\hbar$ soft gravitons of wavelength $r_g=2 G_N M$, where $G_N$ is Newton's constant and $r_g$ corresponds to the Schwarzschild radius. Since the individual gravitational coupling is $\alpha = \hbar G_N/r_g^2$, the system finds itself at the critical point, $\lambda = 1$. An analogous microscopic model exists for the expanding de Sitter spacetime \cite{Dvali:2013eja,Dvali:2017eba}. In this case, the Hubble rate $H$ determines the number of gravitons, $N=1/(\hbar G_N H^2)$, and the gravitational interaction strength, $\alpha=\hbar G_N H^2$. Again we have $\lambda=1$. In summary, criticality emerges as a crucial property of both black holes and de Sitter.

Therefore, one may be able to gain information about gravity by studying our NPM for $\lambda=1$. In this case, we observed $t_q\sim N^\gamma$, with $\gamma \approx 1/2$. Combining this finding with similar results in other systems \cite{Dvali:2015wca,Dvali:2018ytn}, we are tempted to speculate that the quantum break-times for black holes and de Sitter could be
\begin{equation} \label{RootTime}
	t_q \overset{?}{\sim} \sqrt{N} r_g  \;, \qquad \qquad 	t_q \overset{?}{\sim}  \sqrt{N} H^{-1} \;.
\end{equation}
These timescales are much shorter than Page's time \cite{Page:1993wv}, which corresponds to a scaling $t_q\sim N$. If the classical description broke down as early as shown in \eq \eqref{RootTime}, this would have far-reaching consequences for cosmology. Observational constraints on primordial black holes can change radically and small primordial black holes below $10^{15}\,\text{g}$ could become viable dark matter candidates \cite{Dvali:2020wft} (see also \cite{deFreitasPacheco:2023hpb}). Moreover, new observables emerge in the early Universe that are sensitive to the whole history of inflation \cite{Dvali:2018ytn}.

Evidently, it is important to construct analogue models that share more properties with black holes or expanding spacetime than our NPM. Imitating information-processing characteristics \cite{Dvali:2012en,Dvali:2013vxa} appears to be a particularly fruitful avenue; see \cite{Dvali:2020wft} for a concrete proposal. The analogies of gravity with much simpler non-relativistic systems open up an intriguing prospect of answering long-standing questions about black holes and de Sitter in table-top experiments. Analogue models that focus on geometry have already been realized experimentally both for black holes \cite{Lahav:2009wx, Belgiorno:2010wn, Weinfurtner:2010nu} and for cosmology \cite{Eckel:2017uqx,Fifer:2018hcv,Wittemer:2019agm,Steinhauer:2021fhb, Viermann:2022wgw} (see also \cite{Hu:2018psq}). It would be exciting to devise laboratory systems that implement an analogy to gravity based on criticality and information processing.

\section*{Acknowledgments}
We thank Gia Dvali, Rina Kanamoto and Hiroki Saito for extremely useful discussions and insightful feedback. 
We are grateful to  Michael Gedalin and Shira Chapman for granting us access to their HPC resources. 
The work of M.M. was supported by a Minerva Fellowship of the Minerva Stiftung Gesellschaft für die Forschung mbH and in part by the
Israel Science Foundation (grant No. 741/20) and by the German Research Foundation through a German-Israeli Project Cooperation (DIP) grant “Holography and the Swampland”. S.Z. acknowledges support of the Fonds de la Recherche Scientifique - FNRS.

\appendix

\section{Bound on self-coupling}
\label{app:Cm}

In the analytic consideration of section \ref{sec:Model}, we neglected the effect of $C_m$. In the following, we shall study under which conditions this is justified. As before, we shall apply the Bogoliubov-approximation, $\hat{a}_0 \approx \sqrt{N - \Delta N}$, where $\Delta N$ is the number of particles that are not in the $\hat{a}_0$-mode, \ie we immediately took into account backreaction. Up to corrections of order $1/(N-\Delta N)$, we obtain from \eq \eqref{simplifiedSpecies}:
\begin{subequations} \label{simplifiedSpecies2Bogoliubov}
	\begin{align} 
		\hat{H} &\approx \sum_{k=1}^Q \left(\left(1-\frac{1}{2}\lambda \left(1-\frac{\Delta N}{N}\right)\right)\hat{n}_k - \frac{1}{4} \lambda \left(1-\frac{\Delta N}{N}\right) \left(\hat{a}_k^2  +  \hat{a}_k^{\dagger\,2} \right) \right)\\
		& +\frac{C_m}{2}\sum_{k=1}^{Q}\sum_{\substack{l=1\\l\neq k}}^{Q} f(k,l)\left(\hat{a}_{k}^{\dagger\, 2} \hat{a}_{l}^2 +  \text{h.c.}\right) \;.
	\end{align}
\end{subequations}
For the sake of the following consideration, we neglected corrections that scale with $\alpha$ but kept all terms involving $C_m$. Subsequently, we shall use two distinct arguments.

Following a similar reasoning as in \cite{Dvali:2018xoc, Dvali:2020wft}, we can evaluate the expectation value of the Hamiltonian \eqref{simplifiedSpecies2Bogoliubov}. We get approximately
\begin{equation} \label{simplifiedSpeciesBogoliubovExpectation}
	\braket{\hat{H}} = \sum_{k=1}^Q \left(1-\lambda \left(1-\frac{\Delta N}{N}\right)\right) \Delta N + 2 C_m \Delta N \;.
\end{equation}
Here we took into account that due to $f(k,l)$, summing $N_Q^2$ entries with both positive and negative signs leads to a contribution on the order of $N_Q$ (see \eg \cite{Tao2012} and \cite{Dvali:2020wft} where an analogous argument was used).\footnote
{If \eg $\Delta N$ particles are distributed over $Q>\Delta N$ modes with $2$ particles per mode, we get $(\Delta N/2)^2$ non-zero entries and $\braket{\sum_{k=1}^{Q}\sum_{\substack{l=1\\l\neq k}}^{Q} f(k,l)\left(\hat{a}_{k}^{\dagger\, 2} \hat{a}_{l}^2 +  \text{h.c.}\right)} \approx \sqrt{(\Delta N/2)^2} \cdot4 \cdot 2 = 4 \Delta N$.}
Now we impose the requirement that the sign of $\braket{\hat{H}}>0$ should not change because of the inclusion of $C_m$. This amounts to imposing that the additional coupling does not alter the under- or overcriticality of the system. Applying this condition to \eq \eqref{simplifiedSpeciesBogoliubovExpectation} leads to the mild bound
\begin{equation}
	|C_m| \lesssim \frac{1-\lambda}{2} \;.
\end{equation}

Next, we consider a second line of argument, which is only applicable in the undercritical case $\lambda <1$. First, we consider two species, $Q=2$ and then \eq \eqref{simplifiedSpecies2Bogoliubov} gives
\begin{equation}  \label{simplifiedSpecies2Bogoliubov2}
	\hat{H} = \sum_{k=1}^2 \left(\left(1-\frac{1}{2}\lambda \left(1-\frac{\Delta N}{N}\right)\right)\hat{n}_k - \frac{1}{4}\lambda \left(1-\frac{\Delta N}{N}\right) \left(\hat{a}_k^2  +  \hat{a}_k^{\dagger\,2} \right) \right)-\frac{C_m}{2} \left(\hat{a}_{1}^{\dagger\, 2} \hat{a}_{2}^2 +  \text{h.c.}\right) \;,
\end{equation}
where we replaced  the effect of $f(k,l)$ by a negative sign since in this case its effect is maximal, as is evident from the subsequent calculation. Focusing on the $\hat{a}_1$-mode, we will momentarily also assume that $\hat{a}_2$ is macroscopically occupied and approximate $\hat{a}_2\approx \sqrt{\Delta N/2}$, where $N\gg \Delta N \gg 1$.

Then we get
\begin{align}  
	\hat{H}\vert_{k=1} &=  \left(1-\frac{1}{2}\lambda \left(1-\frac{\Delta N}{N}\right)\right)\hat{n}_1 - \frac{1}{2}\left(\frac{1}{2}\lambda \left(1-\frac{\Delta N}{N}\right) + \frac{C_m \Delta N}{2}\right) \left(\hat{a}_1^2  +  \hat{a}_1^{\dagger\,2} \right)\\
	&= \mathcal{E} \hat{n}_1 - \frac{g}{2} \left(\hat{a}_{1}^{2} + \hat{a}_{1}^{\dagger\, 2} \right)  \;,
\end{align}
where we defined
\begin{equation} \label{auxBounds}
	\mathcal{E} = 1 - \frac{1}{2}\lambda \left(1-\frac{\Delta N}{N}\right) \;, \qquad g = \frac{\lambda \left(1-\frac{\Delta N}{N}\right) + C_m \Delta N}{2} \;.
\end{equation}
Correspondingly, we choose the coefficients in the Bogoliubov transformation \eqref{bogoTrafoDefinition} as
\begin{equation} \label{bogoTrafoBounds}
	u^2 = \frac{1}{2}\left(\frac{\mathcal{E}}{\epsilon}+1\right) \;, \qquad
	v^2 = \frac{1}{2} \left(\frac{\mathcal{E}}{\epsilon} -1\right)\;,
\end{equation}
where now
\begin{equation}
	\epsilon = \sqrt{\mathcal{E}^2 - g^2} \;.
\end{equation}
As before, the Hamiltonian becomes
\begin{equation}
	\hat{H} = \epsilon \hat{n}^b \;.
\end{equation}
Plugging the definitions \eqref{auxBounds} in $\epsilon$, we get
\begin{equation}
	\epsilon = \sqrt{\left(1 - \frac{1}{2}\lambda \left(1-\frac{\Delta N}{N}\right)\right)^2 - \left(\frac{\lambda \left(1-\frac{\Delta N}{N}\right) + C_m \Delta N}{2}\right)^2} \;.
\end{equation}
Now we require that the $\hat{a}_1$-mode does not become gapless due to $C_m$. This leads to
\begin{equation}
	|C_m| \lesssim \frac{2(1-\lambda \left(1-\frac{\Delta N}{N}\right))}{\Delta N} \;.
\end{equation}
Finally, we can wonder how this bound change if we couple $\hat{a}_1$ to many hardly-occupied modes $Q$ instead of one highly-occupied one. In this case, we need to take into account the effect of alternating signs in the coupling of the mode. Employing the same argument as in \cite{Dvali:2020wft}, we therefore replace $\Delta N/2$ by $\sqrt{\Delta N}$ to arrive at the bound:
\begin{equation} \label{boundPrel}
	|C_m| \lesssim \frac{1-\lambda \left(1-\frac{\Delta N}{N}\right)}{\sqrt{\Delta N}} \;,
\end{equation}

Minimizing \eq \eqref{boundPrel} with respect to $\Delta N$ and imposing $\Delta N\lesssim N/2$, we get $\Delta N =N \min(1/2, 1/\lambda- 1)$. This leads to\footnote
{In the above reasoning, we did not consider the effect a many unoccupied modes, \ie the case $Q \gg N_Q$. For such parameters, one could replace $\sqrt{N_Q}$ by $\sqrt{Q}$ in \eq \eqref{boundPrel}. Then the sharpest bound would arise close to $\Delta N=0$, which implies $|C_m| \lesssim \frac{1-\lambda}{\sqrt{Q}}$.}
\begin{equation} 
	|C_m| \lesssim \begin{cases}
		\sqrt{3 \alpha} (1-\frac{\lambda}{2}) &  \lambda \leq \frac{2}{3} \\
		2 \sqrt{\alpha(1-\lambda)} & \lambda > \frac{2}{3}
	\end{cases} \;.
\end{equation}
This indicates that $|C_m|$ can be as large as $\sqrt{\alpha}$, provided that $\lambda$ is not too close to $1$. This reasoning, however, is only applicable to the undercritical case $\lambda<1$. Since $\Delta N$ is generically bigger for $\lambda >1$, one could expect that a sharper bound is required such as $|C_m|\lesssim \alpha$.

\section{Undercritical regime}
\label{sec:appendixUnder}

In the following, we show plots containing data as well as fits for the individual scalings, where the main results were already presented in section \ref{sec:numeric}. We start with the undercritical phase defined by $\lambda <1$. Unless stated otherwise, we use \eq \eqref{valuesUndercritical} as base value for all scans in this section and we show the parameter values again for convenience:
\begin{equation} \label{valuesUndercriticalAppendix}
	\lambda = 0.8 \;, \qquad N=50 \;, \qquad C_m = 0.016 \;, \qquad C=4 \;, \qquad Q=10 \;.
\end{equation}
We will use three types of fitting function throughout this and the next sections. These have already been introduced in section \ref{sec:numeric} and we repeat them here for completeness:
\begin{align*}\label{fitAppendix}
	&\text{linear:} \qquad & t_q = m \cdot x + n \;,\\
	&\text{polynomial:} \qquad & t_q = a \cdot x^c + b \;,\\ 
	&\text{logarithm:} \qquad & t_q = p \cdot \log(x) + q \;,\\
	&\text{Shifted polynomial:}  \qquad & t_q = \tilde{a} \cdot (x-\tilde{d})^{\tilde{c}} \;.
\end{align*}

\begin{figure}
	\centering 
	\begin{subfigure}{0.85\textwidth}
		\centering
		\includegraphics[width=0.475\linewidth]{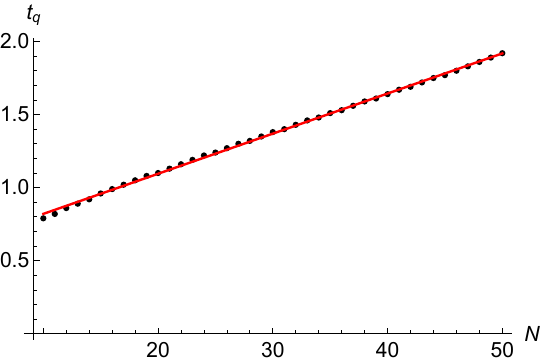}
		\hfill
		\includegraphics[width=0.475\linewidth]{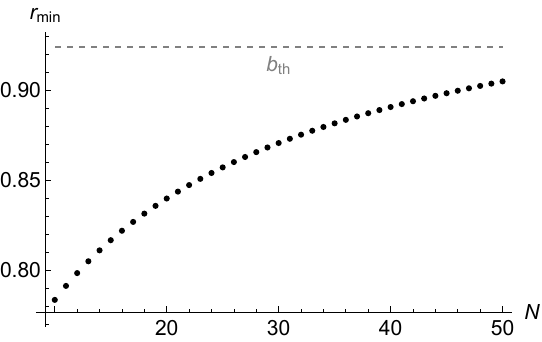}
		\caption{$N$-scan with small $N$.}
		\label{fig:NScanUnder10_50}
	\end{subfigure}
	\vskip\baselineskip
	\begin{subfigure}{0.85\textwidth}
		\centering
		\includegraphics[width=0.475\linewidth]{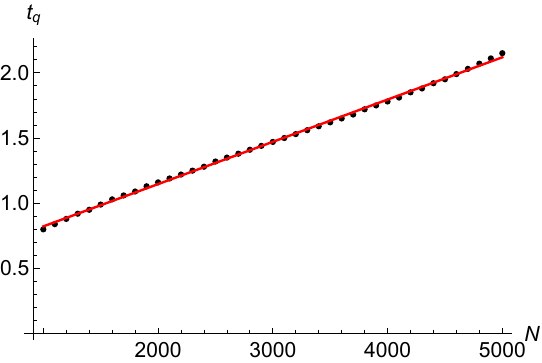}
		\hfill
		\includegraphics[width=0.475\linewidth]{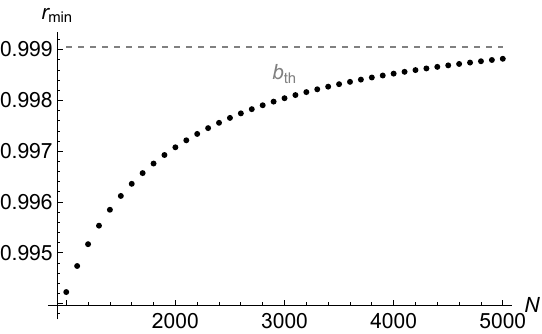}
		\caption{$N$-scan with large $N$.}
		\label{fig:NScanUnder1k_5k}
	\end{subfigure}
	\vskip\baselineskip
	\begin{subfigure}{0.85\textwidth}
		\centering
		\includegraphics[width=0.475\linewidth]{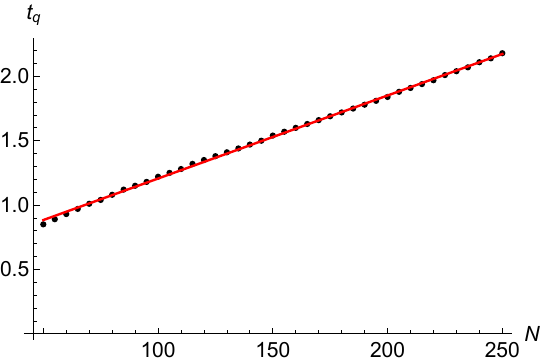}
		\hfill
		\includegraphics[width=0.475\linewidth]{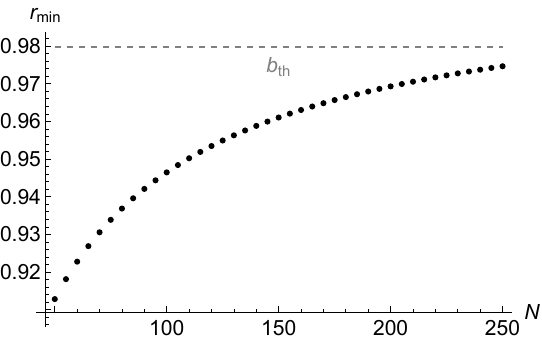}
		\caption{$N$-scan with $C=6$.}
		\label{fig:NScanUnder50_250_C6}
	\end{subfigure}
	\vskip\baselineskip
	\begin{subfigure}{0.85\textwidth}
		\centering
		\includegraphics[width=0.475\linewidth]{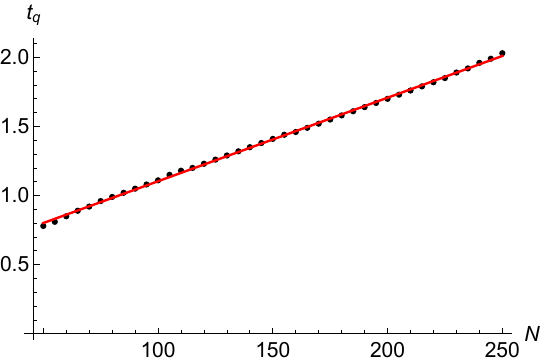}
		\hfill
		\includegraphics[width=0.475\linewidth]{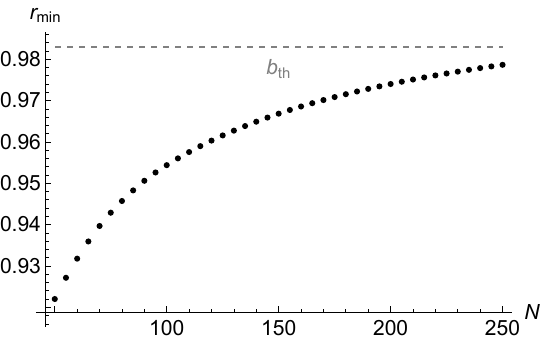}
		\caption{$N$-scan with $C_m=0$}
		\label{fig:NScanUnder50_250_Cm0}
	\end{subfigure}	
	\caption{Various $N$-Scans. (\textbf{Left}) Break-time as a function of total particle number $N$ in the undercritical phase $\lambda<1$. Linear fit depicted in red. Parameters given by \eqref{valuesUndercriticalAppendix}. (\textbf{Right}) Corresponding thresholds (gray dashed line) as well as minimal occupation of the $\hat{a}_0$ mode.}
	\label{fig:NScansUnder}
\end{figure}

\paragraph{Dependence on N}
In addition to the $N$-scan presented in the main text, we perform several others in order to verify the generality of the results. We explained above that the length of a scan interval is limited by the possibility to find an adequate threshold defining the break-time for all values within the range. Fitting over too large interval results in a meaningless definition of quantum breaking. However, it is of course valid to consider subsets of a larger interval and determine a threshold and perform fits on each of those individually. 
Therefore, to show that the scaling \eq \eqref{NlinFitParaUnder} holds generically, we perform similar scans over two additional $N$-ranges. Besides the range presented in the main text, we also scan over a range with smaller particle number $N\in [10,50]$ and one with larger one $N\in [1000,5000]$. We perform a linear as well as a polynomial fit, which yields the following fitting parameters:
\begin{align}
	\text{linear fit}: \qquad	&m = 0.027 \;, \qquad n = 0.55 \;,\nonumber\\
	\text{polynomial fit}:\qquad &a = 0.048\;, \qquad   b = 0.44\;, \qquad  c = 0.87 \;, \nonumber\\
	&    \text{for} \qquad N \in [10,50]\;,\qquad C_m=0.08 \;, 
\end{align}
as well as
\begin{align}
	\text{linear fit}: \qquad	&m = 0.00032 \;, \qquad n = 0.5 \;, \nonumber\\
	\text{polynomial fit}:\qquad &a = 0.0003\;, \qquad  b = 0.5\;, \qquad  c = 1.0 \;,\nonumber\\
	&   \text{for} \qquad N \in [10^3,5\cdot 10^3]\;,\qquad C_m=8 \cdot 10^{-4}\;. \nonumber
\end{align}
Fits for both ranges are depicted in \figs \ref{fig:NScanUnder10_50} and \ref{fig:NScanUnder1k_5k} and we observe a clear linear dependency in both cases. That the smaller $N$-range deviates more from a perfect linear relation can be explained by $1/N$ effects due to the small particle number. Besides that the numerical values for the fit are quantitatively very similar. Apart from the non-trivial threshold fixing, one fitting function could in principle be valid over several orders of magnitude in $N$.

To study the impact of our artificial parameter $C$, we repeat the study from \ref{sec:numericUnder} with a higher capacity $C=6$ over the same set of parameters. The data and fit can be found in \fig \ref{fig:NScanUnder50_250_C6} with fitting values given by
\begin{align}
	\text{linear fit}: \qquad	&m = 0.0065 \;, \qquad n = 0.56 \;, \nonumber\\
	\text{polynomial fit}:\qquad &	a = 0.012\;, \qquad  b = 0.46\;, \qquad  c = 0.90 \;, \nonumber\\    &\text{for} \qquad  C=6 \;.\nonumber
\end{align}

Finally, we also want to take a close look at the influence of $C_m$. In the following, we will set $C_m=0$ and verify that this yields the same result. We vary the total particle number between $50 \le N \le 250$ and apart from setting the interspecies coupling to zero, we keep all other parameters as used in \eq \eqref{valuesUndercriticalAppendix}. This yields
\begin{align}
	\text{linear fit}: \qquad	&m = 0.006 \;,  \qquad n = 0.5 \;, \nonumber\\
	\text{polynomial fit}:\qquad &	a = 0.0074\;, \qquad  b = 0.47\;, \qquad  c = 0.97 \;,\nonumber\\
	&    \text{for} \qquad  C_m=0 \;.
	\nonumber
\end{align}
A exponent $c$ close to one again indicates a linear relation. The corresponding plot is depicted in \fig \ref{fig:NScanUnder50_250_Cm0}.
Summarizing, a linear relation between particle number $N$ and the break-time $t_q$ is observed for all considered parameter choices.

\paragraph{Dependence on Q} 

\begin{figure}
	\centering 
	\begin{subfigure}{0.85\textwidth}
		\centering
		\includegraphics[width=0.49\linewidth]{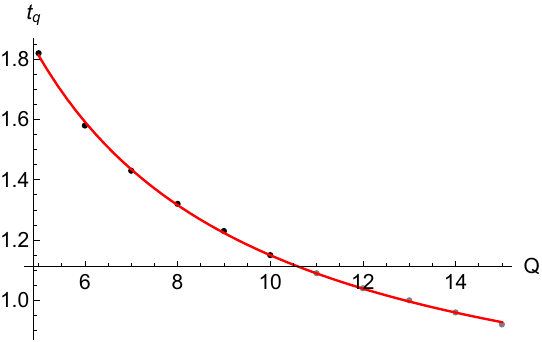}
		\hfill
		\includegraphics[width=0.49\linewidth]{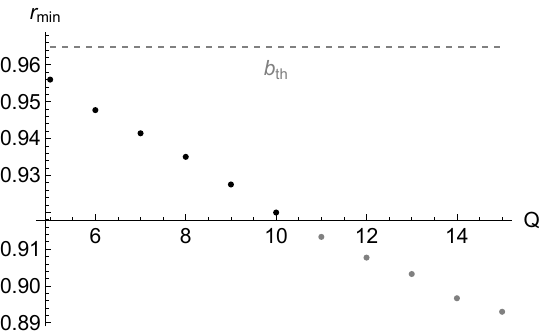}
		\caption{$Q$-scan with $N=50$.}
		\label{fig:QScanUnder}
	\end{subfigure}
	\vskip\baselineskip
	\begin{subfigure}{0.85\textwidth}
		\centering
		\includegraphics[width=0.49\linewidth]{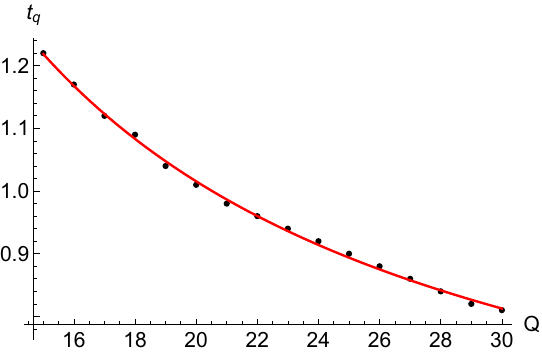}
		\hfill
		\includegraphics[width=0.49\linewidth]{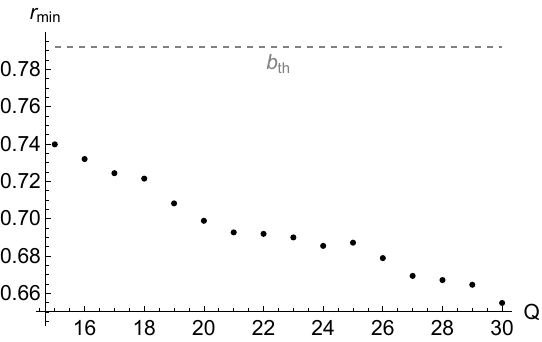}
		\caption{$Q$-scan with $N=10$.}
		\label{fig:QScanUnder2}
	\end{subfigure}
	\caption{Various $Q$-Scans. (\textbf{Left}) Break-time as a function of number of species $Q$ in the undercritical phase $\lambda<1$. Polynomial fit depicted in red. Parameters given by \eqref{valuesUndercriticalAppendix}. (\textbf{Right}) Corresponding thresholds (gray dashed line) as well as minimal occupation of the $\hat{a}_0$ mode.}
	\label{fig:QScansUnder}
\end{figure}

We perform two $Q$-scans. The first one has total occupation number $N=50$ as set by  \eq \eqref{valuesUndercritical} and this scan is also discussed in the main section \ref{sec:numeric}. 
Additionally, we also perform a second one with $N=10$ (and adjusted $C_m$).
The data for both scans are plotted in \figs \ref{fig:QScanUnder} and \ref{fig:QScanUnder2}, respectively. 
A polynomial fit of the form \eq \eqref{polyFit} yields:
	\begin{align}\label{fitvaluesUnderQ}
		&a = 6.57\;, \qquad  \, \, \, b = 0.48\;, \qquad  c = -0.99 \qquad \qquad   \text{for} \qquad N=50,\, \, C_m=0.016 \;. \\
		&a = 12.25\;, \qquad  b = 0.40\;, \qquad  c = -1.00 \qquad \qquad   \text{for} \qquad N=10,\, \, C_m=0.08 \;, \nonumber
	\end{align}
Both values are very close to the analytic expected value of $c=-1$. However, we have to stress that these values vary when changing the fitting range in \fig \ref{fig:QScansUnder}. Especially including smaller $Q$ in the fit leads to an increase in the numerical value of the exponent. This can partly be attributed to the difficulty of finding a meaningful threshold for a large parameter interval. However, some uncertainty seems to exist when fitting $Q$ and extracting numerical fit values.
Apart from this we find that the fitting values, especially the exponent $c$, vary over the same interval independent of the particular $N$. We conclude that the observed $Q$-scaling is stable against the change of the variable $N$.

\begin{figure}
	\centering 
	\begin{subfigure}{0.85\textwidth}
		\centering
		\includegraphics[width=0.49\linewidth]{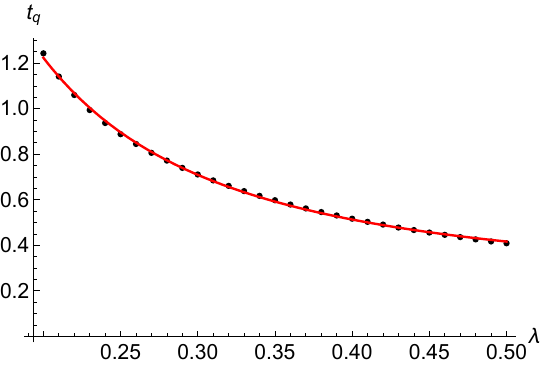}
		\hfill
		\includegraphics[width=0.49\linewidth]{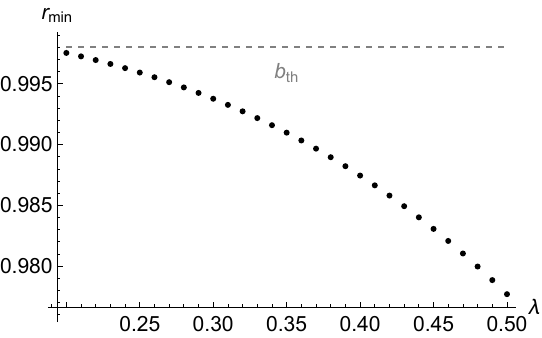}
		\caption{$\lambda$-scan with $0.2 \le \lambda \le 0.5$.}
		\label{fig:LambdaScanUnderSmall}
	\end{subfigure}
	\vskip\baselineskip
	\begin{subfigure}{0.85\textwidth}
		\centering
		\includegraphics[width=0.49\linewidth]{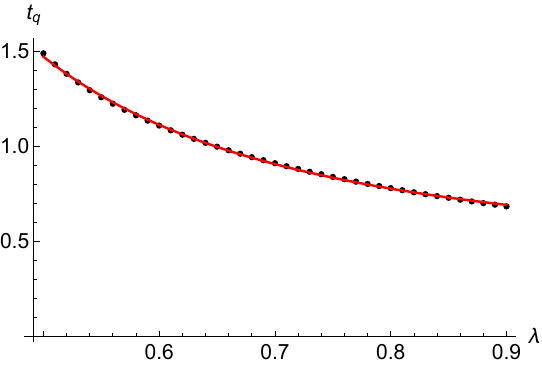}
		\hfill
		\includegraphics[width=0.49\linewidth]{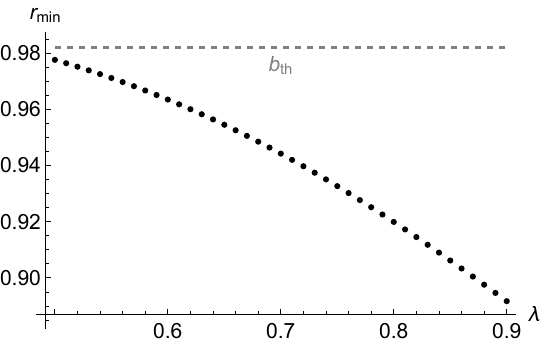}
		\caption{$\lambda$-scan with $0.5 \le \lambda \le 0.9$.}
		\label{fig:LambdaScanUnderLarge}
	\end{subfigure}
	\vskip\baselineskip
	\begin{subfigure}{0.85\textwidth}
		\centering
		\includegraphics[width=0.49\linewidth]{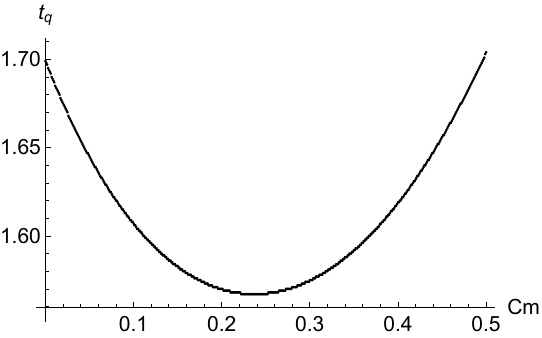}
		\hfill
		\includegraphics[width=0.49\linewidth]{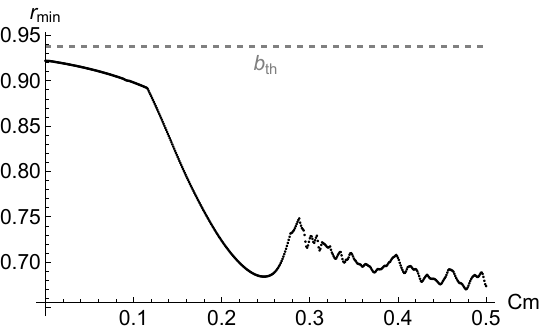}
		\caption{$C_m$-Scan.}
		\label{fig:CmScanUnder}
	\end{subfigure}
	\caption{$\lambda$ and $C_m$-Scans. (\textbf{Left}) Break-time as a function of number of collective coupling $\lambda$ and interspecies coupling strength $C_m$ in the undercritical phase $\lambda<1$. Polynomial fit depicted in red. Parameters given by \eqref{valuesUndercriticalAppendix}. (\textbf{Right}) Corresponding thresholds (gray dashed line) as well as minimal occupation of the $\hat{a}_0$ mode. }
	\label{fig:ScansUnder}
\end{figure}

\paragraph{Dependence on $\lambda$}
Starting from the base values \eqref{valuesUndercritical}, we vary the collective coupling in this paragraph within the range $0.2\le \lambda \le 0.9 $. However, as mentioned several times by now, fitting over too large intervals is problematic due to inadequate thresholds. For this reason we split the interval in two subintervals and perform a fit on each section separately. The first one is defined by $0.2 \le \lambda \le 0.5$, the second one by $0.5 \le \lambda \le 0.9$. The data and fit can be found in \fig \ref{fig:LambdaScanUnderSmall} and \fig \ref{fig:LambdaScanUnderLarge} respectively. The detail fit parameters are given by
	\begin{align}\label{fitvaluesUnderLambda}
		&a = 0.059\;, \qquad  b = 0.22\;, \qquad  c =-1.76 \qquad \qquad   \text{for} \qquad 0.2 \le \lambda \le 0.5 \;, \\
		&a = 0.22\;, \qquad \, \, \,   b = 0.41\;, \qquad  c = -2.26 \qquad \qquad   \text{for} \qquad 0.5 \le \lambda \le 0.9 \;. \nonumber
	\end{align}	
Again the exponent $c$ shows fluctuations when fitting over different values. This can probably again be traced to threshold effects. The fitting parameters for the exponent on the two different intervals seem to spread around the analytical estimated scaling of $t_{q} \sim 1/\lambda^2$.

\paragraph{Dependence on $C_m$}
As explained in the main text, we do not expect $C_m$ to play a major role over the value range that we consider. To check this expectation on the concrete system, we again perform numerical simulation while varying the interspecies coupling over the range $0 \le C_m \le 0.5$. Again we use \eq \eqref{valuesUndercritical} as base values for the parameters of the system. The data is plotted in \fig \ref{fig:CmScanUnder}. It is clear from the plot that there is only a marginal dependence of the break-time on $C_m$ despite the fact we vary $C_m$ to much larger values than we consider in the remainder of this paper. We therefore do not perform a fit. It is however interesting to note that in the undercritical phase there exist two different regimes. In the first one, increasing $C_m$ leads to a reduction of the break-time while going beyond a certain value ($C_m\sim 0.24$ in this specific instance) the break-time increases gain and departure from the initial state takes a longer period of time. 

\section{Overcritical regime with few species}
\label{sec:appendixOverFew}
Next we show data and fits for the overcritical phase with $\lambda>1$. This phase can be categorized in two regimes according to section \ref{sec:Model}. We follow the same order as in the main part and start with the case when $Q$ is small compared to the number of total particles $N$, corresponding to the first case in \eq \eqref{breaktimeOvercriticalSpecies}. Unless stated otherwise, all scans in this section take 
\eq \eqref{valuesOvercriticalFew} as their base values. which we print again for convenience:
\begin{equation} \label{valuesOvercriticalFewAppendix}
	\lambda = 1.2 \;, \qquad N=50 \;, \qquad C_m = 0.016 \;, \qquad C=50 \;, \qquad Q=3 \;.
\end{equation}

\begin{figure}
	\centering 
	\begin{subfigure}{0.85\textwidth}
		\centering
		\includegraphics[width=0.49\linewidth]{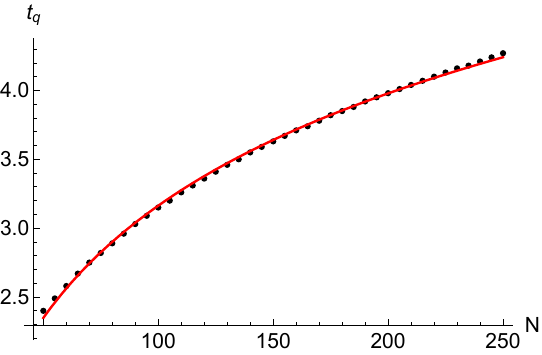}
		\hfill
		\includegraphics[width=0.49\linewidth]{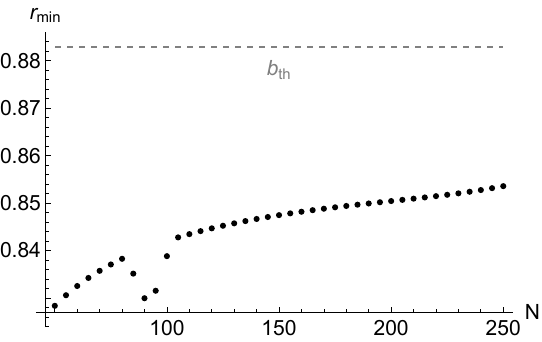}
		\caption{$N$-Scan with $C_m=0$.}
		\label{fig:NScanOverFewCm0}
	\end{subfigure}
	\vskip\baselineskip
	\begin{subfigure}{0.85\textwidth}
		\centering
		\includegraphics[width=0.49\linewidth]{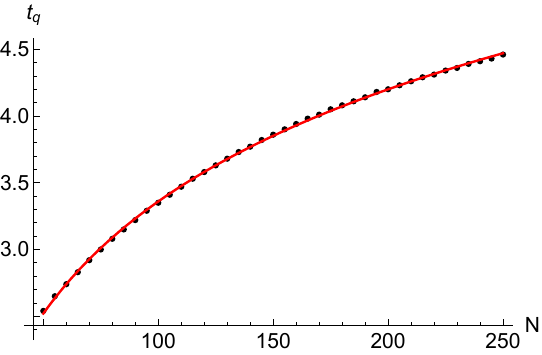}
		\hfill
		\includegraphics[width=0.49\linewidth]{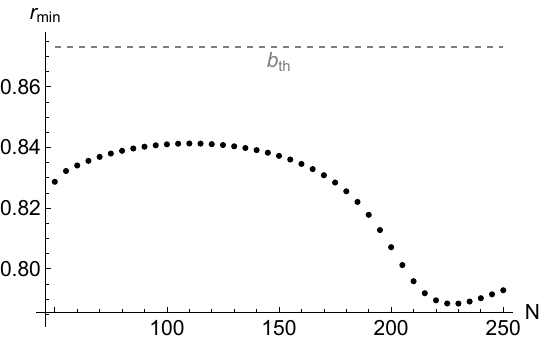}
		\caption{$N$-Scan with $C=60$.}
		\label{fig:NScanOverFewC60}
	\end{subfigure}
	\caption{Various $N$-Scans. (\textbf{Left}) Break-time as a function of total particle number $N$ in the overcritical phase $\lambda>1$ with few species. Logarithmic fit depicted in red. Parameters given by \eqref{valuesOvercriticalFewAppendix}. (\textbf{Right}) Corresponding thresholds (gray dashed line) as well as minimal occupation of the $\hat{a}_0$ mode.}
	\label{fig:NScansOverFew}
\end{figure}

\paragraph{Dependence on N}
The main scan about the $N$-dependency of the break time is in full detail presented in the main text in section \ref{sec:numericOverFew}. In addition to the one shown there, we perform here a simulation with higher capacity $C=60$ and vanishing interspecies coupling $C_m=0$. Besides these changes, the scan range $50 \le N \le 250$ as well as the remaining parameters (see \eq \eqref{valuesOvercriticalFew}) are kept the same. The data and fit for higher capacity can be found in \fig \ref{fig:NScanOverFewC60}, those for $C_m=0$ are presented in \fig \ref{fig:NScanOverFewCm0}. Fitting a logarithmic function of the form \eq \eqref{logfit} to both sets yields the following parameters:
\begin{align}\label{fitvaluesOverNAdd}
	&p=1.18\;, \qquad  q = -2.25 \qquad \qquad   \text{for} \qquad C_m=0 \;, \\
	&p = 1.21\;, \qquad  q = -2.22 \qquad \qquad   \text{for} \qquad C=60 \;. \nonumber
\end{align}	
In all three sets (base, higher $C$, vanishing $C_m$), the values  for the fit parameters  are very similar and verify the robustness of our numerical findings. Notice, however, although the fits are very similar the precise break-times are different.

\begin{figure}
	\centering 
	\begin{subfigure}{0.85\textwidth}
		\centering
		\includegraphics[width=0.49\linewidth]{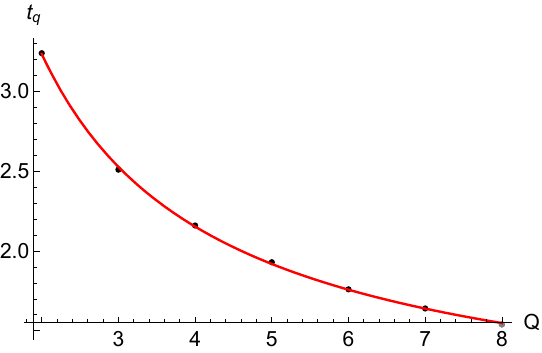}
		\hfill
		\includegraphics[width=0.49\linewidth]{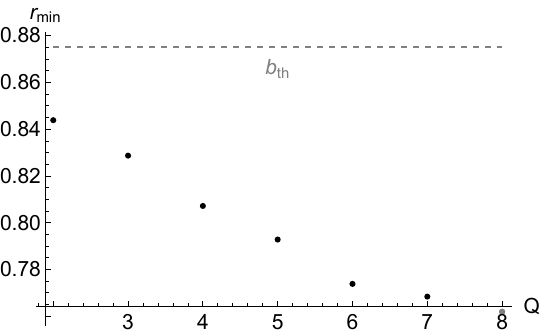}
		\caption{$Q$-Scan with a polynomial fit function \eqref{polyFit}.}
		\label{fig:QScanOverFew}
	\end{subfigure}
	\vskip\baselineskip
	\begin{subfigure}{0.85\textwidth}
		\centering
		\includegraphics[width=0.49\linewidth]{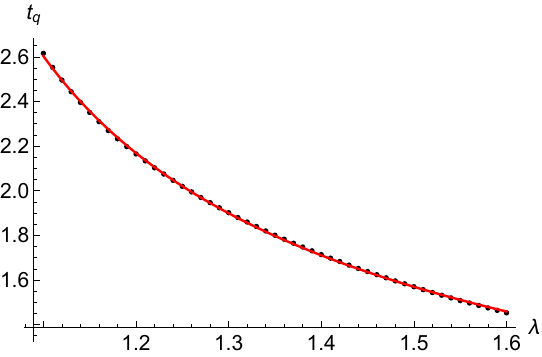}
		\hfill
		\includegraphics[width=0.49\linewidth]{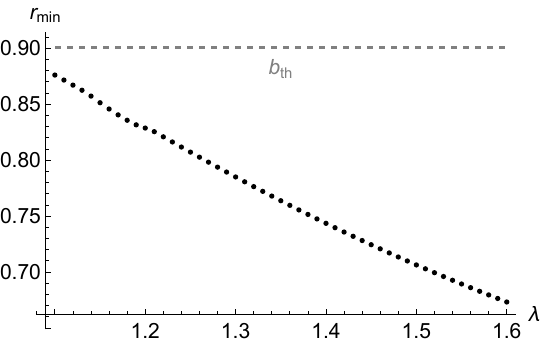}
		\caption{$\lambda$-Scan with a shifted fit function \eqref{fitShiftedPoly}.}
		\label{fig:lambdaScanOverFew2}
	\end{subfigure}
	\vskip\baselineskip
	\begin{subfigure}{0.85\textwidth}
		\centering
		\includegraphics[width=0.49\linewidth]{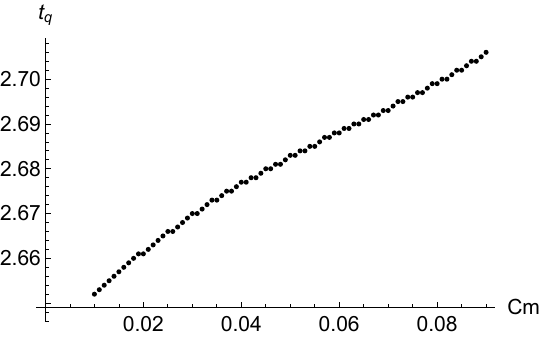}
		\hfill
		\includegraphics[width=0.49\linewidth]{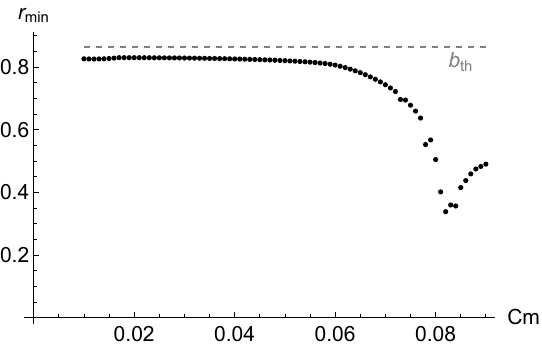}
		\caption{$C_m$-Scan.}
		\label{fig:CmScanOverFew}
	\end{subfigure}
	\caption{$Q, \lambda$ and $C_m$-Scans. (\textbf{Left}) Break-time as a function of number of the amount of species $Q$, the collective coupling $\lambda$ and interspecies coupling strength $C_m$ in the overcritical phase $\lambda>1$ with few species. (Shifted) polynomial fit depicted in red. Parameters given by \eqref{valuesOvercriticalFewAppendix}. (\textbf{Right}) Corresponding thresholds (gray dashed line) as well as minimal occupation of the $\hat{a}_0$ mode. }
	\label{fig:ScansOverFew}
\end{figure}

\paragraph{Dependence on Q} 
We run simulation with varying the species number within $2\le Q\le 8$. The data, which is concisely summarized in section \ref{sec:numericOverFew}, can be reviewed in more detail in \fig \ref{fig:QScanOverFew}. Moreover, the precise fitting values are given by
\begin{equation}
	a =  4.39\;, \qquad  b = 0.78\;, \qquad  c = -0.84 \;.
\end{equation}
Again we observer a minor deviation from the analytic expectation $t_q \sim 1/Q$. We have to emphasize that the precise values for the above fit depend on the range one fits. In summary, although the precise numerical value are model dependent, they still seem near the analytic expectation within numerical uncertainties. 

\paragraph{Dependence on $\lambda$}
The data for the $\lambda$ scan presented in the main text can be found in \fig \ref{fig:lambdaScanOverFew2}. As discussed previously, it can be described very accurately by a shifted polynomial function defined in \eq \eqref{fitShiftedPoly}. The fit yields following values: 
\begin{equation} \label{fitValuesOverFewLambda}
		\tilde{a} =1.24\,, \hspace{20pt} \tilde{d} =  0.88 \, , \hspace{20pt} \tilde{c} = -0.49 \;.
\end{equation}
The data and the fit are visualized in \fig \ref{fig:lambdaScanOverFew2}.

\paragraph{Dependence on $C_m$}
Completely analogously to the undercritical phase, the break-time should not depend on $C_m$ significantly. We test this claim again by scanning of $0.01 \le C_m \le 0.09$. The data can be found in \fig \ref{fig:CmScanOverFew}.
As expected, there is only a marginal influence of $C_m$ on the break-time in the range of values considered. Note that the threshold determined by the smallest amplitude in this scan seems inadequate for the dynamics with high amplitudes for larger coupling constant $C_m \gtrapprox 0.08$. However, the qualitative behavior seems to be unaffected by this choice, as it is evident from the continuous and slow change in \fig \ref{fig:CmScanOverFew}.

\section{Overcritical regime with many species}
\label{sec:appendixOverMany}
Next, we present the data and details about the fits of section \ref{sec:numericOverMany}. This regime lies in the overcritical phase $\lambda >1$ with a large number of species as compared to the total occupation number of the system. Specifically, it corresponds to the second line in \eq \eqref{breaktimeOvercriticalSpecies}.
As base values, we use  \eq \eqref{valuesOvercriticalMany} unless states otherwise and we print them again for convenience:
\begin{equation} \label{valuesOvercriticalManyAppendix}
	\lambda = 1.2 \;, \qquad N=10 \;, \qquad C_m = 0.08 \;, \qquad C=16 \;, \qquad Q=10 \;.
\end{equation}

\begin{figure}
	\centering 
	\begin{subfigure}{0.85\textwidth}
		\centering
		\includegraphics[width=0.49\linewidth]{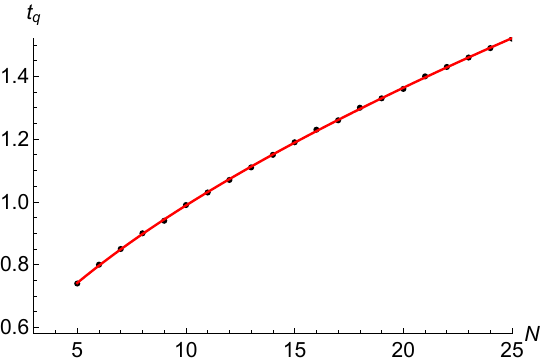}
		\hfill
		\includegraphics[width=0.49\linewidth]{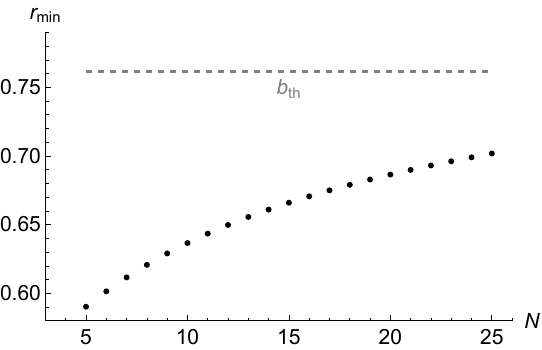}
		\caption{$N$-Scan with $C_m=0$.}
		\label{fig:NScanOverManyCm}
	\end{subfigure}
	\vskip\baselineskip
	\begin{subfigure}{0.85\textwidth}
		\centering
		\includegraphics[width=0.49\linewidth]{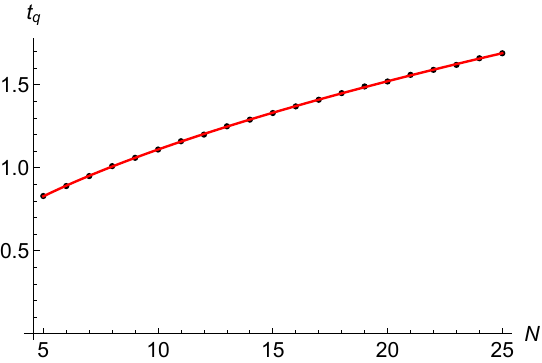}
		\hfill
		\includegraphics[width=0.49\linewidth]{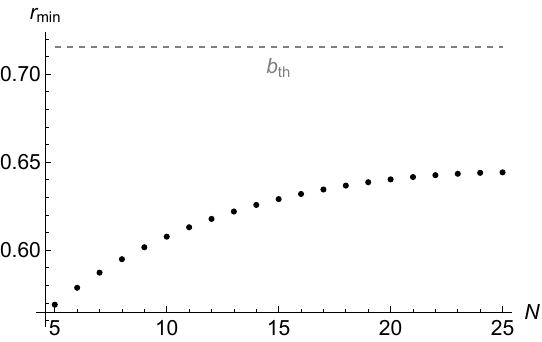}
		\caption{$N$-Scan with $C=20$.}
		\label{fig:NScanOverManyC}
	\end{subfigure}
	\caption{Various $N$-Scans. (\textbf{Left}) Break-time as a function of total particle number $N$ in the overcritical phase $\lambda>1$ with many species. Polynomial fit depicted in red. Parameters given by \eqref{valuesOvercriticalManyAppendix}. (\textbf{Right}) Corresponding thresholds (gray dashed line) as well as minimal occupation of the $\hat{a}_0$ mode.}
	\label{fig:NScansOverMany}
\end{figure}

\paragraph{Dependence on N}
Similarly to the other overcritical regime, we check the generality of the result presented in the main section \ref{sec:numericOverMany}. Therefore, we again perform two additional scans with vanishing interspecies coupling $C_m$ as well as higher capacity $C$. Again, we vary $N$ within the interval $5 \le N \le 25$. The data and corresponding polynomial fits can be reviewed in \figs \ref{fig:NScanOverManyCm} and \ref{fig:NScanOverManyC}, respectively. The fit parameters have the following numerical values
\begin{align}\label{fitvaluesOverManyNAdd}
	&a = 0.18\;, \qquad  b = 0.27 \;,   \qquad  c = 0.60 \qquad \qquad   \text{for} \qquad C_m=0.0 \;, \\
	&a = 0.26\;, \qquad  b = 0.21\;, \qquad  c = 0.54 \qquad \qquad   \text{for} \qquad C=20 \;. \nonumber
\end{align}
In analogy to the other $N$-scans, the corresponding scalings are insensitive to the $C$ (as long as it is chosen large enough) as well as the precise value of $C_m$. Specifically, the exponent of the polynomial seems to spread around the numerical value $c=0.5$. Note that $N$ gets very small in these scans so we expect that finite size effects can have an influence in this analysis.

\begin{figure}
	\centering 
	\begin{subfigure}{0.85\textwidth}
		\centering
		\includegraphics[width=0.49\linewidth]{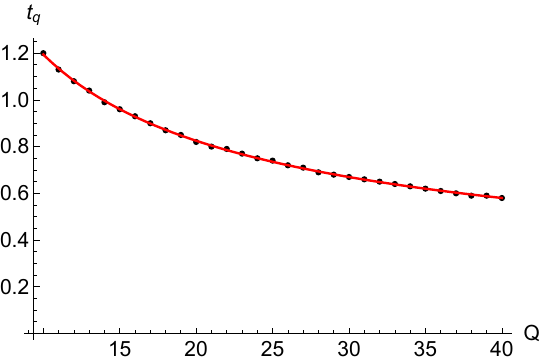}
		\hfill
		\includegraphics[width=0.49\linewidth]{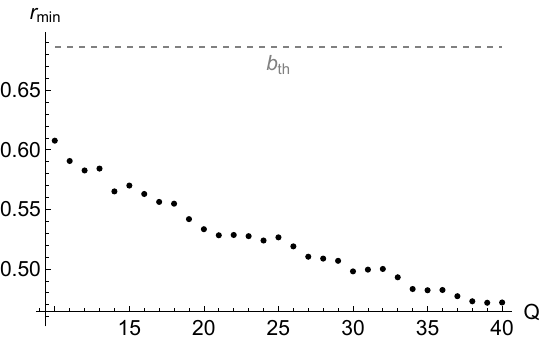}
		\caption{$Q$-Scan with $N=10$ and $C_m=0.08$.}
		\label{fig:QScanOverMany}
	\end{subfigure}
	\vskip\baselineskip
	\begin{subfigure}{0.85\textwidth}
		\centering
		\includegraphics[width=0.49\linewidth]{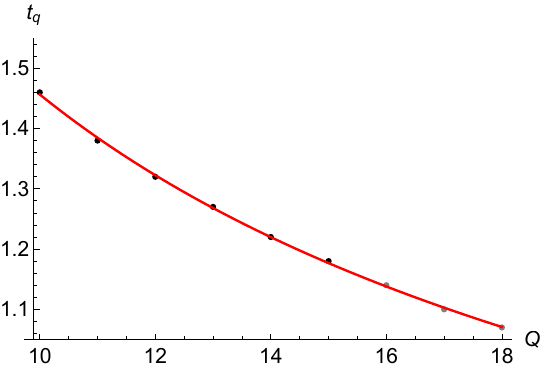}
		\hfill
		\includegraphics[width=0.49\linewidth]{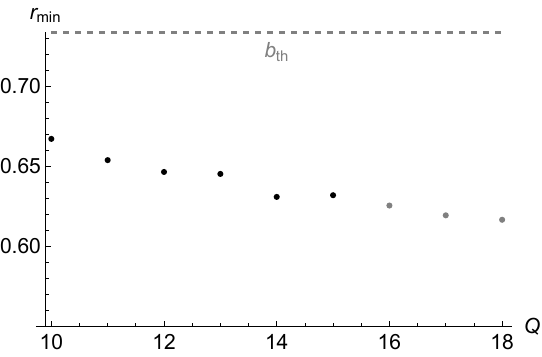}
		\caption{$Q$-Scan with $N=20$ and  $C_m=0.04$.}
		\label{fig:QScanOverMany3}
	\end{subfigure}
	\vskip\baselineskip
	\begin{subfigure}{0.85\textwidth}
		\centering
		\includegraphics[width=0.49\linewidth]{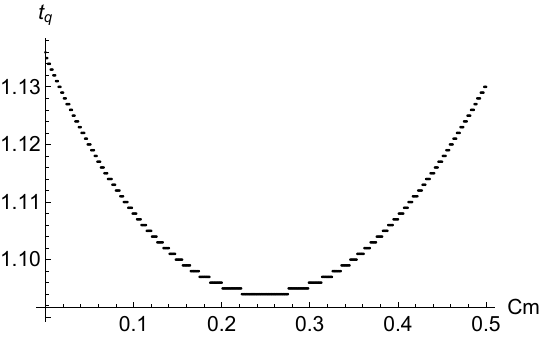}
		\hfill
		\includegraphics[width=0.49\linewidth]{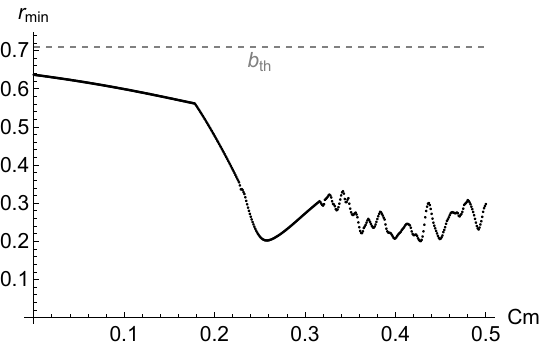}
		\caption{$C_m$-Scan.}
		\label{fig:CmScanOverManyW}
	\end{subfigure}
	\caption{$Q$ and $C_m$-Scans. (\textbf{Left}) Break-time as a function of number of the amount of species $Q$ and interspecies coupling strength $C_m$ in the overcritical phase $\lambda>1$ with many species. (Shifted) polynomial fit depicted in red. Parameters given by \eqref{valuesOvercriticalManyAppendix}. (\textbf{Right}) Corresponding thresholds (gray dashed line) as well as minimal occupation of the $\hat{a}_0$ mode. }
	\label{fig:ScansOverMany}
\end{figure}

\paragraph{Dependence on Q} 
In order to remain in the many species regime of the overcritical phase, we have to choose $Q$ large enough compared to $N$ according to \eq \eqref{breaktimeOvercriticalSpecies}. For this reason we will vary $Q$ within $10 \le Q \le 40$ in the following. The fit and data presented in the main section can be seen in \fig \ref{fig:QScanOverMany}. In addition to the scan defined by \eq \eqref{valuesUndercritical} and presented in section \ref{sec:numericOverMany}, we also run a scan with higher total occupation and correspondingly adjusted $C_m$. Due to limited computational resources, increasing the total particle number $N$ requires to reduce the scan range to $10 \le Q \le 18$. We also adjust the minimal $Q$ to compensate for the larger $N$ to stay in the many species regime. The data and fit  can be seen in \fig \ref{fig:QScanOverMany3}. For both runs, we summarize the fit parameters in the following table
\begin{align}\label{fitvaluesOverManyQAdd}
	&a = 4.29\;, \qquad  b = 0.098\;, \, \, \, \, \, \, \, \, \,   c = -0.59 \qquad  \qquad   \text{for} \qquad N=10, \, \, C_m=0.08 \;, \\
	&a = 5.07\;, \qquad  b = 0.12\;, \qquad  c = -0.58 \qquad \qquad   \text{for} \qquad N=20, \, \, C_m=0.04 \;. \nonumber
\end{align}

\begin{figure}
	\centering 
	\begin{subfigure}{0.85\textwidth}
		\centering
		\includegraphics[width=0.49\linewidth]{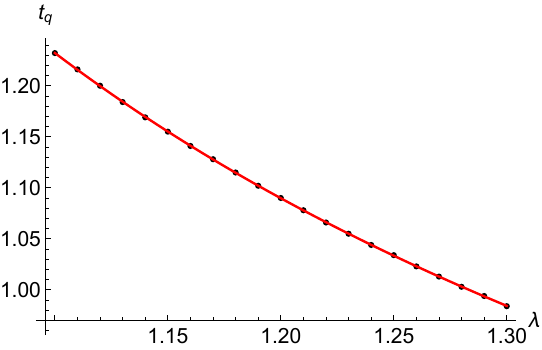}
		\hfill
		\includegraphics[width=0.49\linewidth]{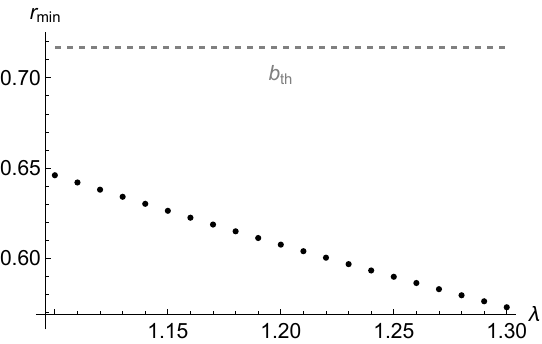}
		\caption{$\lambda$-Scan with fit function \eq \eqref{fitShiftedPoly}.}
		\label{fig:lambdaScanOverMany}
	\end{subfigure}
	\vskip\baselineskip
		\begin{subfigure}{0.85\textwidth}
		\centering
		\includegraphics[width=0.49\linewidth]{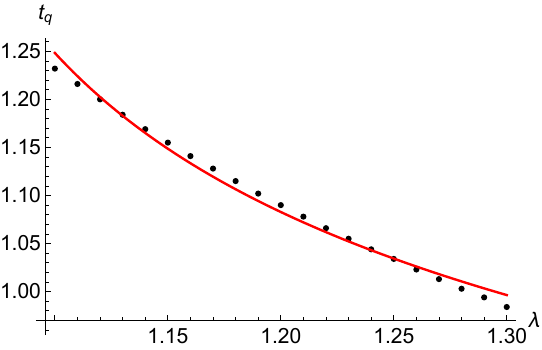}
		\hfill
		\includegraphics[width=0.49\linewidth]{omLthres.pdf}
		\caption{$\lambda$-Scan with fit function \eq \eqref{fitShiftedPoly} and fixed $d=1$.}
		\label{fig:lambdaScanOverManyFixedd}
	\end{subfigure}
	\vskip\baselineskip
	\begin{subfigure}{0.85\textwidth}
	\centering
	\includegraphics[width=0.49\linewidth]{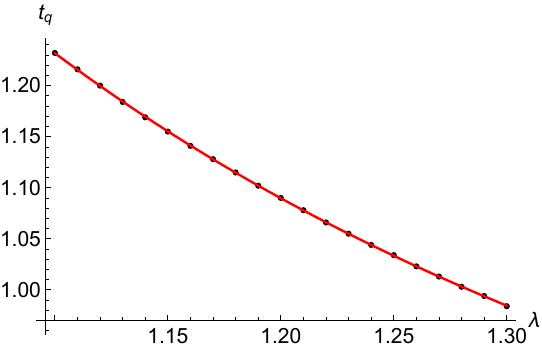}
	\hfill
	\includegraphics[width=0.49\linewidth]{omLthres.pdf}
	\caption{$\lambda$-Scan with polynomial fit function \eq \eqref{polyFit}.}
	\label{fig:lambdaScanOverManyPoly}
\end{subfigure}
	\caption{$\lambda$-Scan with different fit functions. (\textbf{Left}) Breaktime as a function of number of the amount of the collective coupling $\lambda$ in the overcritical phase $\lambda>1$ with many species. Fits depicted in red. Parameters given by \eqref{valuesOvercriticalManyAppendix}. (\textbf{Right}) Corresponding thresholds (gray dashed line) as well as minimal occupation of the $\hat{a}_0$ mode.}
	\label{fig:ScansOverManyLambda}
\end{figure}

\paragraph{Dependence on $\lambda$}
As illustrated in \eq \eqref{summaryScalings} we were not able to obtain a conclusive analytic understanding of the scaling of the break-time with the collective coupling $\lambda$ in the overcritical phase in the presence of many accessible modes. We therefore perform several different fits over the interval $1.1 \le \lambda \le 1.3$. 
The obtained data with a shifted polynomial fit can be seen in \fig \ref{fig:lambdaScanOverMany}. Motivated by the known critical point of quantum phase transition $\lambda =1$ in the limit of large $N$ we also perform a fit with fixed shift $\tilde{d}=1$ in \fig \ref{fig:lambdaScanOverManyFixedd}. We note that fixing the shift seems to worsen the fit. Additionally, we also show the same data with a polynomial fit in \fig \ref{fig:lambdaScanOverManyPoly}.
The numerical values for the fit function are given by
\begin{align}
	&\tilde{a} = 0.75\,, \hspace{20pt} \tilde{d} =  0.66 \, , \hspace{20pt} \tilde{c} = -0.60 \qquad \qquad   \text{for function \eq \eqref{fitShiftedPoly}\;,} \nonumber\\
	&\tilde{a} = 0.78 \, , \hspace{20pt} \tilde{c} = -0.21 \qquad \qquad   \text{for function \eq \eqref{fitShiftedPoly} and fixed $\tilde{d}=1$\;,}  \\
	&a = 0.90\;, \qquad  b = 0.53\;, \qquad  c = -2.59 \qquad \qquad   \text{for function \eq \eqref{polyFit}\;.} \nonumber
\end{align}
The value differs more significantly from the square root indication found in the few species regime for the same fit function \eqref{polyFit}.  We emphasize, however, that a polynomial fit with exponent $c=-2.6$ describes the data equally accurate and it is at this point not clear which (or another) scaling is the true dependency on $\lambda$.

\paragraph{Dependence on $C_m$}
Last but not least, we perform a $C_m$-scan over the range $0 \le C_m \le 0.5$ with parameter choice \eqref{valuesOvercriticalMany}.
The data is visualized in \fig \ref{fig:CmScanOverManyW}. Similarly to the undercritical phase discussed earlier, also this regime shows no significant change over the whole interval. We therefore conclude that $C_m$ is unimportant for the break-time in our prototype system \eq \eqref{simplifiedSpecies}.

\section{Critical interpolation}
\label{sec:appendixCritInterpolation}

\begin{figure}
	\centering 
	\begin{subfigure}{1.0\textwidth}
		\centering
		\includegraphics[scale=1.1]{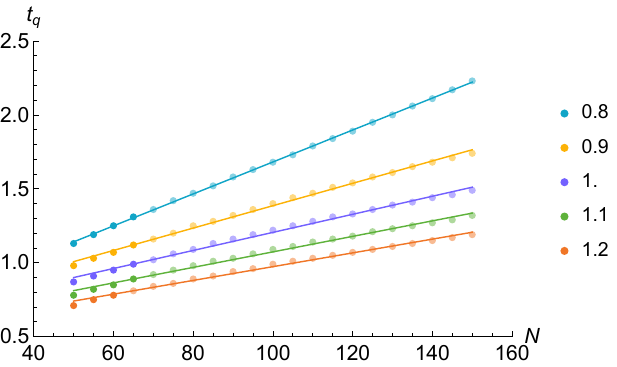}
		\caption{Linear fits.}
		\label{fig:InterpolationLinCThres}
	\end{subfigure}
	\hspace{0.05\textwidth}
	\begin{subfigure}{1.0\textwidth}
		\centering
		\includegraphics[scale=1.1]{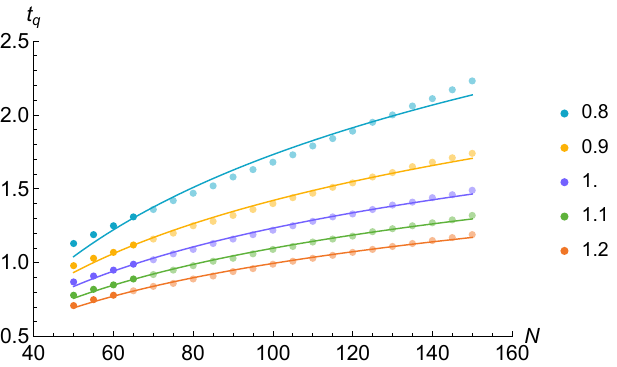}
		\caption{Logarithmic fits.}
		\label{fig:InterpolationLogCThres}
	\end{subfigure}
	\caption{Quantum break-time as a function of $N$ for various $\lambda$ interpolating between the undercritical regime, $\lambda <1$, and the overcritical phase, $\lambda >1$, where a single common threshold for defining quantum breaking is used in all fits (unlike in \fig \ref{fig:Interpolation}). Linear fits in \fig \ref{fig:InterpolationLinCThres} and logarithmic fits in \fig \ref{fig:InterpolationLogCThres} on the same data. Low collective coupling $\lambda$ are described by a linear relation, a $\log N$ dependence describes the scaling more accurately for higher $\lambda$. Increasing the collective coupling $\lambda$ leads to a faster break-time.}
	\label{fig:InterpolationCommonThres}
\end{figure}

We performed several $N$-scans in section \ref{sec:numericInterpolation} to interpolate between the undercritical and overcritical (with few species) phase of system \eqref{simplifiedSpecies}. As discussed above, the dynamics changes continuously from a linear scaling to logarithmic one. The break-time as a function of $N$ increases for larger $\lambda$ in \fig \ref{fig:Interpolation}. This is not in contradiction with various $\lambda$-scans  throughout this paper (see for example \fig \ref{fig:lambdaScanOverFew2} in section \ref{sec:numericOverFew}). As already explained, this is an artifact of choosing a separate threshold for each $\lambda$ individually. As a last consistency check, we perform on the same data a similar analysis, however, this time we choose a single threshold for all time evolutions. The result is shown in \fig \ref{fig:InterpolationCommonThres} and now a larger collective coupling leads to a faster break-time, in accordance with previous findings. Also here a linear fit works best for $\lambda < 1$ while the data can be more accurately described by a $\log$ fit for $\lambda >1 $ with a continuous transformation in between. However, the distinction is significantly less clear. The reason for this is that a common threshold, which is set by the smallest amplitude in the set, is unsuitable for cases in which a large maximal amplitude of $n_0/N$ is reached.

\clearpage

\bibliographystyle{JHEP}
\bibliography{QB}
	
\end{document}